\documentclass[oldversion,twocolumn]{aa}
\usepackage{ifpdf}
\usepackage{graphicx,amssymb,lineno}

\newcommand{\indis}[1]{{\mbox{\scriptsize #1}}}

\input epsf

\begin{document}

\author{I.V.~Panov\inst{1,2} \and H.-Th.~Janka\inst{1}}

\title{On the dynamics of proto-neutron star winds
and r-process nucleosynthesis}   


%
%
%
\offprints{I.V.~Panov,
\email{Igor.Panov@itep.ru}}
%
\institute{Max-Planck-Institut f\"ur Astrophysik,
Karl-Schwarzschild-Stra\ss e 1, D-85748 Garching, Germany
 \and
  Institute for Theoretical and Experimental
 Physics,  B.~Cheremushkinskaya 25, Moscow, 117259, Russia.
 }


\abstract{
We study here the formation of heavy r-process nuclei in the
high-entropy environment of rapidly expanding neutrino-driven
winds from compact objects. In particular,
we explore the sensitivity of the element creation in the
$A \ga 130$ region
to the low-temperature behavior of the outflows. For this purpose
we employ a simplified model of the dynamics and of the
thermodynamical evolution for radiation dominated, adiabatic
outflows. It consists of a first stage of fast, exponential
cooling with timescale $\tau_\mathrm{dyn}$, followed
by a second phase of slower evolution, assuming either constant
density and temperature or a power-law decay of these quantities.
These cases describe a strong deceleration or
decreasing acceleration of the transsonic outflows, respectively,
and thus are supposed to capture the most relevant effects associated
with a change of the wind expansion behavior at large radii,
for example because of the
collision with the slower, preceding supernova ejecta and the
possible presence of a wind termination shock.
We find that for given entropy, expansion timescale, and
proton-to-baryon ratio not only the transition temperature between
the two expansion phases can make a big difference in the formation
of the platinum peak, but also the detailed cooling law during the
later phase. Because the nuclear photodisintegration rates
between about $2\times 10^8\,$K and roughly $10^9\,$K are
more sensitive to the temperature than the neutron-capture rates
are to the free neutron density, a faster cooling but continuing
high neutron density in this temperature
regime allow the r-process path to move closer to the neutron-drip
line. With low ($\gamma$,n)- but high $\beta$-decay rates,
the r-processing does then not proceed through a
($\gamma$,n)-(n,$\gamma$) equilibrium but through a quasi-equilibrium
of (n,$\gamma$)-reactions and $\beta$-decays, as recently also
pointed out by Wanajo. Unless the transition temperature and
corresponding (free neutron) density become too small
($T \la 2\times 10^8\,$K), a lower temperature or faster temperature
decline during the slow, late evolution
phase therefore allow for a stronger appearance of the
third abundance peak.

}


\titlerunning{Proto-neutron star winds and r-process nucleosynthesis}

\authorrunning{Panov \& Janka}

\maketitle

\section{Introduction}

Heavy nuclei beyond the iron peak are known to be produced in nature
mainly through neutron capture reactions (Burbidge   et al.  1957).
Rapid neutron capture and the reverse photodisintegration processes
achieve an equilibrium among the isotopes of each heavy element. Then
beta-decay occurs that leads to the increase of the nuclear charge and
formation of a new element. When the neutron capture rate is much higher than
the beta-decay rate ($\lambda_{\mathrm{n}\gamma} > \lambda_{\beta}$) and
$T_9\sim 1$ ($T_9$ is defined as the temperature normalized to $10^9\,$K),
the r-process can start and, for a sufficiently high
neutron-to-seed ratio, the wave of nucleosynthesis drives the process to
heavier nuclei, forming, in part, the abundance r-process peaks at
$\sim$80, 130 and 196.

Although the r-process sites remain unknown, many astrophysical models
and sources for r-process elements
have been proposed during the last 50 years,
including, in particular, such scenarios as that of an explosion on a
neutron-star surface (Bisnovatyi-Kogan and Chechetkin 1979), a
collision of a neutron star with a black hole (Lattimer \& Schramm,
1976), an explosion of a low-mass neutron star (Imshennik, 1992), and
the hypothetical escape of nucleon bubbles in case of a
soundless stellar collapse (Imshennik \& Litvinova 2006). The latter
scenario was suggested as origin of gamma-ray bursts but could also
be an interesting site for the nucleosynthesis of heavy elements.

Presently, however, it seems most likely that rapid n-capture nucleosynthesis
can take place during different stages of supernova explosions
(e.g., Hillebrandt  1978; Woosley \& Hoffman  1992), or in
neutron star mergers (e.g., Lattimer \& Schramm  1974;
Symbalisty \& Schramm 1982; Freiburghaus et al. 1999). An overview of
the currently discussed possible sites is given in a recent review
paper by Arnould et al.\ (2007).

Supernovae and neutron star mergers have different advantages and weak
points, but the main difference lies probably in the initial
neutron-to-proton ratio, which is necessary to support a sufficiently
high free neutron density during several hundreds of milliseconds. With
respect to this parameter, all astrophysical scenarios can be separated
into two distinct groups, in which nucleosynthesis can be carried out
over short or long timescales (so-called short-time or long-time
solutions, respectively), as introduced by Seeger et al.\ (1965).
In nature, the r-process might be realized in both these types of
sources and also other ones might contribute (see for example Cameron,
2003; Arnould et al.\ 2007). It is only future modelling and
observations that will be able clarify the multiplicity of the
r-process models and types, as well as the contributions from different
production sites to the r-process element abundances observed in the
Solar System.

Among the  astrophysical events proposed as sites for the r-process,
supernova explosions still remain the preferable ones (see, e.g.,
Wanajo \& Ishimaru 2006, and references therein). In particular,
supernova explosions can distribute r-process material all over the
Galaxy, and estimated amounts of heavy elements produced in SN
explosions are in accordance with the observations.


%
The neutrino-driven wind from a hot neutron star produced in a supernova
explosion has been considered as a probable site for the r-process by many
authors (see {\em e.g.} Meyer   et al.\ 1992; Woosley   et al.\ 1994;
Witti et al.\ 1993; Otsuki et al.\ 2000; Sumiyoshi et al.\ 2000;
Terasawa et al.\ 2001; Wanajo et al.\ 2001).
A part of the surface material of a neutron star is
heated by the supernova neutrinos and gets ejected. It can be described as
a hot outflow with a fairly high entropy and a moderate density.

Various studies of the r-process element formation in proto-neutron
star winds  have been conducted during the past years. Hoffman, Woosley
\& Qian (1997), employing the analytic wind model of Qian \& Woosley
(1996), explored the possibility of third peak creation for different
combinations of the determining parameters of $Y_\indis{e}$ (the
electron-to-baryon ratio), entropy, and expansion timescale.
Altogether, they showed (see their Fig.~10) that for typical values of
$Y_\mathrm{e}\ga 0.4$
in the wind the third r-process peak can be produced for combinations
ranging from moderate entropy ($s \sim 100$ in units of Boltzmann's
constant per baryon) and very short timescales ($t_\mathrm{exp}$ a few
milliseconds) to high entropies ($s \ga 400$) and long expansion timescales
(a few 100 milliseconds). The wind models existing at that time failed
to provide the necessary conditions (Witti et al.\ 1994, Qian \& Woosley
1996).

Subsequently, several other studies of neutrino-driven winds in the
framework of general relativistic gravity (Otsuki et al.\ 2000,
Sumiyoshi et al.\ 2000, Thompson et al.\ 2001, Wanajo et al.\ 2001,
2002) confirmed the need of fairly extreme conditions concerning
expansion timescale or entropy for strong r-processing up to $A \sim
200$. Since very high entropies could not be obtained in the wind
scenario or were found to be associated with too low mass loss rates
for any significant production of r-nuclei, these studies also
demonstrated a preference for the case of moderate entropies, $s \sim
100$--200, and very short timescales. This seemed to give a bias for
winds from compact neutron stars with a large mass, $M \ga 2\,M_\odot$,
and a small radius $R \la 10\,$km as the most likely site for r-process
element formation up to the platinum peak.

This conclusion was found to hold independent of whether
freely expanding, transsonic ``wind'' outflows were considered
(Hoffman et al.\ 1997, Thompson et al.\ 2001) or subsonic
``breeze'' solutions (Otsuki et al.\ 2000;
Sumiyoshi et al.\ 2000; Wanajo et al.\ 2001;
Terasawa et al.\ 2001, 2002) or supersonic winds with some
fixed freeze-out temperature $T_\mathrm{f}$ (Wanajo et al.\ 2002,
Wanajo 2007). Winds
are characterized by a monotonically increasing velocity and a
continuous temperature decrease when the radius goes to infinity,
whereas breezes
are obtained when a prescribed, non-vanishing value of the
pressure and temperature is required to be reached at the outer
boundary at some large radius
(this is supposed to mimic the fact that the fast neutrino-driven
wind is decelerated again as it collides and merges with the
preceding, more slowly expanding material ejected in the early phase
of the supernova blast). The approach taken by Wanajo et al.\ (2002)
allowed them to causally disconnect the conditions at large radii from
those at the neutron star and thus to choose the asymptotic temperature
independent of the mass-loss rate (in contrast to the situation for
breezes and winds). The results from these different types
of outflow scenarios concerning combinations of entropy, $Y_\mathrm{e}$,
and expansion timescale necessary for strong r-processing
in the ejecta turned out to be in qualitative
agreement (compare, e.g., Fig.~11 in Thompson et al.\ 2001 with
Fig.~8 in Otsuki et al.\ 2000 and with Table~2 and Fig.~6 in
Sumiyoshi et al.\ 2000).

Later Terasawa et al.\ (2002) announced to have found successful
conditions even for proto-neutron stars with a more typical mass around
1.4$\,M_\odot$ and a radius of 10$\,$km. Like previous studies (in
particular Sumiyoshi et al.\ 2000, but also Otsuki et al.\ 2000) they
considered breeze outflows, but different from the earlier
investigations they chose a smaller value for the outer boundary
pressure, which implied a lower asymptotic temperature. They argued
that this is favorable for a strong r-processing up to the third peak
because the lower final temperature is associated with a faster
expansion and more rapid cooling. The more quickly decreasing
temperature leads to a slowing down of charged-particle reactions and
reduces the efficiency of $\alpha$-particle recombination. Therefore it
leads to less production of seed nuclei and a higher neutron-to-seed
ratio. Such a sensitivity to the expansion timescale was also seen by
Arnould et al.\ (2007), who performed systematic variations of the
parameters in analytic wind and breeze solutions. Arnould et al.\
(2007) verified that wind solutions provide the more favorable
conditions for strong r-processing than the slower breezes, because for
given values of the entropy and mass loss rate the expansion timescale
is directly correlated with the asymptotic value of the temperature and
thus of the total specific energy of the outflowing gas. These values
are lowest in case of freely expanding winds. Arnould et al.\ (2007),
however, also saw that the mass loss rate has a much more sensitive
influence: breeze solutions with higher mass loss rates (and otherwise
the same characterising parameter values) make a faster expansion and
allow for a stronger r-process despite having higher asymptotic values
of the temperature.

Although Arnould et al.\ (2007) confirmed the formation of the third
r-process peak for combinations of entropy, expansion timescale,
$Y_\mathrm{e}$, neutron star mass, and asymptotic temperature
in the ballpark of those considered by Terasawa et al.\ (2002),
the results of the latter paper are nevertheless in contradiction to
the earlier studies by Otsuki et al.\ (2000), Sumiyoshi et al.\ (2000),
and Thompson et al.\ (2001):
Terasawa et al.\ (2002) obtained a significantly higher entropy and
shorter expansion timescale than Sumiyoshi et al.\ (2000) even for
the same choice of outflow determining conditions, i.e., for
the same individual neutrino luminosity
($L_{\nu_i} = 10^{51}\,$erg$\,$s$^{-1}$), the same mean neutrino energies,
and in particular the same outer boundary pressure (compare Table~1 in
the Terasawa et al.\ work and Section~3.2 in Sumiyoshi et al.).
Moreover, for all tested values of the boundary
pressure, Terasawa et al.\ (2002) found outflow properties
for their 1.4$\,M_\odot$ neutron star that were
largely different from those plotted for breezes
in Fig.~8 of Otsuki et al.\ (2000)
and for freely expanding winds in Fig.~11 of Thompson et al.\ (2001).
These differences seem to have been causal for the successful solar system
like r-process reported by Terasawa et al.\ (2002), but the actual
reason why the more favorable outflow behavior was obtained, remains
unexplained\footnote{Note that Thompson et al.\ (2001) used a
sophisticated description of the equation of state and thus
explicitly accounted
for the nonrelativistic character of electrons and positrons at
low temperatures, an effect that Sumiyoshi et al.\ (2000)
pointed out to be important
for a reliable determination of the expansion timescale.}.

Recently, Arcones et al.\ (2007) performed new hydrodynamic simulations
of neutrino-driven winds, systematically exploring the effects of the
wind termination shock that forms when the supersonic wind collides
with the slower earlier supernova ejecta (Janka \& M\"uller 1995a,b;
Burrows et al.\ 2005; Buras et al.\ 2006, see also Tom{\`a}s et al.\
2004) and that decelerates the outflow abruptly. They found that the
position of the reverse shock is strongly dependent on the evolution
phase, progenitor structure, and explosion energy of the supernova.
Motivated by these studies, interest has recently turned to a closer
exploration of the relevance of the late-time wind dynamics for
r-process nucleosynthesis. On the one hand, Wanajo (2007)   showed
numerically   that a solar-like r-process can also be produced in
supersonically expanding outflows whose temperature drops quickly to a
few $10^8\,$K instead of asymptoting to a value around $10^9\,$K as
previously mostly assumed (Otsuki et al.\ 2000, Sumiyoshi et al.\ 2000,
Wanajo et al.\ 2001, 2002). In such a low-temperature environment an
$(\mathrm{n},\gamma)$-$(\gamma,\mathrm{n})$ equilibrium is never
achieved during the nucleosynthesis of heavy r-process material, but
neutron captures compete with $\beta$-decays in the low-density matter,
similar to what was discussed by Blake \& Schramm (1976).

On the other hand, Kuroda et al.\ (2008) started to explore for the
first time systematically the consequences of the wind termination
shock for the r-processing in the wind. Decelerating the outflow
abruptly, the reverse shock does not only raise the entropy of the
matter, but in particular it slows down the temperature and density
decline that takes place in the subsequent expansion. Kuroda et al.\
(2008)   pointed out   that the change of the temperature behavior
plays a decisive role in determining the r-process abundances, because
the nucleosynthesis path depends strongly on the temperature during the
r-process freeze-out phase. In contrast, the entropy jump does not seem
to be important because high entropies in the shocked outflows are
reached only when the temperature is already well below $2.5\times
10^9\,$K, i.e.\ not in the regime where the neutron-to-seed ratio is
established before the onset of r-processing.

In the present paper we also investigate the influence
of the late-time outflow dynamics on the r-process
nucleosynthesis. To this end we consider outflow trajectories that
consist of an initial homologous phase and a second, slower expansion
stage. This describes the wind dynamics only schematically but at least
some basic features that are found in detailed solutions of
unshocked as well as shocked outflows can be reproduced.
In contrast to the outflow solutions studied
by Arnould et al.\ (2007), for example, our simple parametric
ansatz for the temperature and density decrease in the ejected
matter allows us to modify the early expansion behavior and that
at late times independently, i.e.\ in an uncoupled way.
This is similar to the approach taken by Wanajo et al.\ (2002)
and closer to how, for example,
the presence of a wind termination shock
acts on the matter. Motivated by the insensitivity to the entropy
jump seen by Kuroda et al.\ (2008), we ignore the discontinuity
of the fluid variables at the location where their
time dependence is assumed to change.

In order to obtain r-processing with reasonable instead of
extreme values of the
wind entropy $s$ and electron fraction $Y_\mathrm{e}$, we consider
sufficiently short exponential timescales for the first expansion
stage, i.e., we consider timescales in the ballpark of those given
for example by Otsuki et al.\ (2000) for outflows from
a neutron star with a gravitational mass of 2$\,M_\odot$, a
radius of 10$\,$km, and a neutrino luminosity of $10^{52}\,$erg$\,$s$^{-1}$
(see their Table~1). We will vary the prescribed
expansion behavior during the second phase in order to discuss the
effects of the late-time temperature evolution on the r-processing
from a more general point of view than done in previous works.
The role of a strong
or weak deceleration of the very fast, transsonic winds will be
investigated by a constant or power-law behavior, respectively, of
the temperature and density during the second stage. The former
may be interpreted as the limiting case of a strong deceleration
by a wind termination shock. The latter may be considered as an
approximation to the reduced acceleration that a freely expanding
wind experiences at larger radii after its roughly homologous
initial phase.

Our paper is structured as follows.
In Sect.~2 we will describe the dynamical model and the network
used in our study, and will present some numerical tests
we performed. In Sect.~3 we will describe our results and in
Sect.~4 we finish with conclusions.

\section{Network, input data, and numerical modeling of the
($\alpha$+r)-process}
\label{sec:input}

We consider here the conditions for r-process nucleosynthesis
in neutrino-driven outflows from the surface layers of hot nascent
neutron stars, which have been the subject of many previous studies,
where different aspects of the problem were discussed
(see, e.g., Qian \& Woosley 1996; Hoffman et al. 1997; Woosley et
al. 1994; Cardall \& Fuller  1997; Qian \& Wasserburg  2000; Takahashi
et al.  1994; Otsuki   et al.  2000; Sumiyoshi   et al.  2000; Thompson
  et al.  2001; Kuroda et al.\ 2008, Arnould et al.\ 2007, and references
therein). Our main goal here is to evaluate numerically the possibility
of producing r-process elements in this environment in dependence of
the late-time behavior of the outflowing gas. Within the framework of a
very simple, purely analytic model   of the wind    we want to
determine the favorable combinations of entropy, $Y_\indis{e}$ and
dynamical timescale, and how sensitively these parameters influence the
outcome.    Though, of course, our results will not allow us to make a
judgement about whether neutrino-driven winds are the long-sought site
of r-process material or not,  our parametric approach has the
advantage of reducing the dynamical aspects of the problem to an
absolute minimum of ingredients, giving one much freedom in the choice
of the involved parameter values.

\subsection{Outflow behavior}
\label{sec:outflow}

We represent the outflow behavior during the early and late expansion
phases by different analytic functions. These describe qualitatively
(but certainly not to very high accuracy) on the one hand
the wind acceleration through the sonic point, and on the other hand
the evolution of the outflow during a second phase of either deceleration
by a reverse shock or reduced acceleration.

In the first stage of the expansion of a spherical mass shell we
assume a homologous velocity-radius dependence, $v \propto r$,
corresponding to an exponential growth of the radius,
$r(t) = R_{\mathrm{ini}}\exp(t/\tau_{\mathrm{dyn}})$. Therefore,
steady-state conditions (which imply $r^2\rho v\,=\,$const)
yield an exponential decline of the density
and for an adiabatically expanding, radiation-dominated wind
(i.e., $\rho\propto T^3$) also an exponential decrease of the
temperature:
\begin{eqnarray}
\rho(t) &=& \rho_\indis{ini} \cdot \exp(-3t/\tau_\indis{dyn})\, , \\
T_9(t) &=& T_9^{\mathrm{ini}} \cdot \exp(-t/\tau_\indis{dyn})\, .
\label{ur1}
\end{eqnarray}
Here $T_9$ is the temperature normalized to $10^9\,$K and
$\rho_{\mathrm{ini}}$ and $T_9^{\mathrm{ini}}$ are the initial
values of density and temperature at some small radius $R_{\mathrm{ini}}$.
The dynamical timescale or expansion timescale $\tau_{\mathrm{dyn}}$
will be treated as a free parameter and can vary between
1$\,$ms and more than 100$\,$ms.

For reasons of simplicity, we will always assume that $\rho$ is
proportional to $T^3$ and that the gas entropy per nucleon
(in units of Boltzmann's constant $k_{\mathrm{B}}$) is given by
the relation $s = 3.34\,T_9^3/\rho_5$ with
$\rho_5 = \rho/(10^5\,$g$\,$cm$^{-3}$). One should, however,
keep in mind that this is only a simplifying approximation,
which is accurate only when the entropy of the gas is large
enough ($s \ga 100\,k_{\mathrm{B}}$ per nucleon) and the temperature
sufficiently high ($T_9 \ga 5$). In this case radiation-dominated
conditions prevail, electrons are relativistic, electron-positron
pairs are abundant, and baryons contribute to the total entropy
only at the level of a few percent. In reality, however, the pairs
begin to disappear below $T_9\approx 5$ and the ratio $T^3/\rho$
increases. Assuming it to be constant also at low temperatures
therefore leads to an overestimation of the density in the
outflowing matter compared to a truely adiabatic evolution
(see Witti et al.\ (1994) for a detailed discussion).

The deceleration by the reverse shock or changing
acceleration behavior are supposed to happen at a
radius $r_0$ and time $t_0$ when the velocity, density, and
temperature reach the values $v_0$, $\rho_0$, and $T_0$, respectively.
As discussed in the introduction, we ignore here the discontinuous
behavior of the dynamical and thermodynamical variables at the
shock and connect the early-time behavior continuously with the
late-time behavior. This means that also the ratio $T^3/\rho$
(and therefore the quantity $s$ that we consider as gas entropy)
remains unchanged at the transition point and
is taken to be constant during the following second stage of the
expansion. For the latter we consider two
cases with different limiting behavior for $t \gg t_0$. In the first
case we assume that the density and temperature asymptote to
constant values,
\begin{eqnarray}
\rho(t)& =& \rho_0\,,\label{eq:late0} \\
T(t) &=& T_0 \,,\ \mathrm{for}\ t\,\geq\,t_0\,.
\label{eq:late1}
\end{eqnarray}
For steady-state conditions this implies that the radius and the
velocity of a Lagrangian mass shell evolve at $t \geq t_0$ according to
\begin{eqnarray}
r(t) &=& r_0\,\left[ 1+ 3\,{v_0\over r_0}\,(t-t_0)\right]^{1/3}\,,
\label{eq:late0r} \\
v(t) &=& v_0\,\left[ 1+ 3\,{v_0\over r_0}\,(t-t_0)\right]^{-2/3}\,,
\label{eq:late1r}
\end{eqnarray}
and therefore $r(t)\propto t^{1/3}$ and
$v(t) \propto t^{-2/3} \propto r^{-2} \to 0$
for $t \gg t_0$. We consider this choice of time-dependence as
an approximative representation of the strong deceleration experienced
by the wind in passing through a termination shock.
Assuming constant density and
temperature on the typical timescale of r-processing (several hundred
milliseconds) is in reasonably good agreement with the slow evolution
of the shocked outflow found in detailed hydrodynamic models (see
Figs.~6 and 8 in Arcones et al.\ 2007).

In the second
investigated case the density and temperature are still assumed
to decline at late times, but much less steeply than during
the exponential first expansion phase:
\begin{eqnarray}
\rho(t)& =& \rho_0\,\left({t\over t_0}\right)^{\! -2}\,, \label{eq:late2}\\
T(t) &=& T_0\,\left({t\over t_0}\right)^{\! -2/3}\,,\
\mathrm{for}\ t\,\geq\,t_0\,.
\label{eq:late3}
\end{eqnarray}
For steady-state conditions this corresponds to
\begin{eqnarray}
r(t) &=& r_0\,\left[ 1 - {v_0t_0\over r_0} + {v_0t_0\over r_0}
\left({t\over t_0}\right)^{\! 3}\right]^{1/3}\,, \label{eq:late2r} \\
v(t) &=& v_0\,\left[ 1 - {v_0t_0\over r_0} + {v_0t_0\over r_0}
\left({t\over t_0}\right)^{\! 3}\right]^{-2/3}
\!\! \left({t\over t_0}\right)^{\! 2}\,,
\label{eq:late3r}
\end{eqnarray}
which yields $r(t)\propto t$ and
$v(t) = v_0^{1/3}(r_0/t_0)^{2/3} =\mathrm{const} >0$ for $t \gg t_0$.
A posteriori this asymptotic behavior justifies our
choice of the power-law time-dependence for $\rho$ and $T$ in
Eqs.~(\ref{eq:late2}) and (\ref{eq:late3}).
Note that $v(t\to \infty) < v_0$, i.e.\ deceleration (and not just
a slow-down of the acceleration) happens only at late times
if $t_0 > \tau_\mathrm{dyn}$.
Values of the parameters used for some of the considered model
cases are listed in Table~\ref{tab:parameters}.

The slower decline of the density and temperature during the
power-law phase may be considered as a simplified representation of
a reduced wind acceleration occurring after an
approximately homologous (i.e., $v \propto r$) early expansion.
Neutrino-driven winds
exhibit such a nearly homologous evolution and thus nearly exponential
increase of the radius and velocity with time only up to some
distance, but then the velocity continues to grow less rapidly even
in the absence (or before) a possible strong deceleration by a
termination shock.
The basic properties of this behavior are captured by our approach.

\begin{table*}[ht]
\caption{           
 Parameter values for some of our considered outflows and of
 relevance for third peak formation.}
\begin{tabular}{|r|rrr|rcc|cc|}
 \hline
run &  $\tau_\indis{dyn}$ & entropy & $T_9^{\mathrm{f}}(t_0)$  &
$t_0$ & $v_\indis{ini}$ & $v_0$  &
$\left\langle A\right\rangle $  & $\left\langle Y(A_3)/Y(A_2)\right\rangle $ \\
  &  [ms] & [$k_\mathrm{B}$/N] &   & [ms] &
[km/s] & [km/s]  &  &  \\
\hline
1 &  5    &  105 & 1   &  $\phantom{\int^0}$9
                              & 2000   & 12000 &  118  & $10^{-10}$ \\
2 &  2.5  &  105 & 1   &  4.5 & 4000   & 24000 &  127  & 0.01       \\
3 &  5    &  145 & 1.4 &  8   & 2000   & 10000 &  128  & 0.075      \\
4 &  5    &  145 & 0.4 & 14   & 2000   & 33000 &  128  & 0.04       \\
5 &  2.5  &  145 & 1   &  4   & 4000   & 20000 &  144  & 0.45       \\
6 &  10   &  170 & 1   & 16   & 1000   &  5000 &  119  & $10^{-5}$  \\
7 &  5    &  170 & 1   &  8   & 2000   & 10000 &  145  & 0.75       \\
\hline
\end{tabular}
\label{tab:parameters}
\end{table*}

\subsection{Calculating the nucleosynthesis}
\label{sec:nucleosynthesis}

At the onset of its expansion, the ejected matter in the neutrino-driven
wind is very hot and composed of free neutrons and protons. With ongoing
cooling the nuclear statistical equilibrium (NSE) shifts towards an increasing
mass fraction of alpha particles until finally the recombination to
heavy nuclei sets in. Provided the conditions of entropy, electron fraction,
and expansion timescale are suitable, the production of elements between
the first and second peaks of the abundance curve through
charged particle reactions and neutron captures may occur. For sufficiently
extreme conditions even third-peak elements and ($A\sim 196$) may be
assembled. Typically starting our nucleosynthesis calculations at
temperatures $T_9 \sim 6$ and densities
$\rho \sim 10^5$--$10^6\,$g$\,$cm$^{-3}$ (the initial density is 
chosen according to a specified value for the conserved gas entropy), 
the element formation in our
model runs first proceeds mainly through charged-particle reactions.
As the temperature and density decrease, the importance of
($\alpha$,X)- and inverse reactions diminishes and the nuclear flow
begins to be driven by a dynamical r-process, provided a sufficient
number of free neutrons is still present.
The production of a small mass fraction of heavy nuclei with $Z \geq 26$
before the onset of rapid neutron captures significantly reduces the
requirements for the neutron source (a smaller number of free
neutrons is required).

The time-dependent concentrations of nuclear species, $Y(A,Z)$, during
the r-processing as determined by reactions with neutrons,
beta-decays, and fission processes are described by the nucleosynthesis
network developed by Blinnikov \& Panov (1996) and Nadyozhin et al.\
(1998). This network was here extended by charged-particle reactions
and a larger number of nuclear reactions and fission processes so that
it is possible to handle both the initial $\alpha$-process and the
subsequent r-process with the same code.

The number of nuclei and reaction equations included in calculations
depends on boundary conditions and the employed nuclear mass model
and can be as large as about 4300. We considered the
region of nuclei with $Z$ ranging from 3 to 100. The minimum and maximum
atomic mass values for each chemical element were determined by the proton
and neutron drip lines.

The reaction rates entering the system of differential equations
differ by tens of orders of magnitude.
Thus, the system of equations for nuclear kinetics
to be solved is a classical example of a stiff system of
ordinary differential equations. We used one of the most effective
methods to integrate such a stiff system of equations, Gear's method
(Gear, 1971). The description of the complete package of solver
routines and its applications to the r-process calculations can be
found in Nadyozhin et al.\ (1998).


Nuclear mass values as predicted by FRDM (M\"oller et al.\ 1995)
were used, the beta-decay rates were calculated in the framework of the
QRPA-model (Kratz et al.\ 1993), and the reaction rates with neutrons
were described according to the calculations of Cowan et al.\ (1991)
and of Rauscher \& Thielemann (2000).


In the nucleosynthesis studies presented here, the triple $\alpha$ and
$\alpha\alpha{\mathrm n}$ reactions of helium burning, $3\alpha
\rightarrow ^{12}$C and $\alpha + \alpha + {\mathrm n} \rightarrow
^9$Be, respectively, along with their inverse reactions, were included.
The rates for both processes  were taken from the REACLIB library
of Thielemann et al.\ (1987) and those for the reactions of heavier
nuclei with protons and
$\alpha$-particles from Rauscher \& Thielemann (2000).

\subsection{Code tests and comparisons}
\label{sec:tests}

Code tests for a number of different cases were performed by
\cite{pbt01}, and for a number of explored cases r-process
calculations with the same rates gave
practically the same isotopic yields as in the paper of \cite{frt99}.
Network calculations of the $\alpha$-process with the present code were
compared with the results shown by Witti et al.\ (\cite{wjt94})
and yield rather compatible abundance distributions at the
beginning of the r-process, in spite
of differences in the employed nuclear reaction rates and the nuclear
mass model.

In this context we would like to note that the transition from the
$\alpha$-process to the r-process can only be done correctly on the
basis of the same mathematical model. In our calculations we use the
same code for both parts of the nucleosynthetic reaction sequence,
without artificial devision into $\alpha$- and r-process steps.

%
%
\begin{figure*}[tpb!]
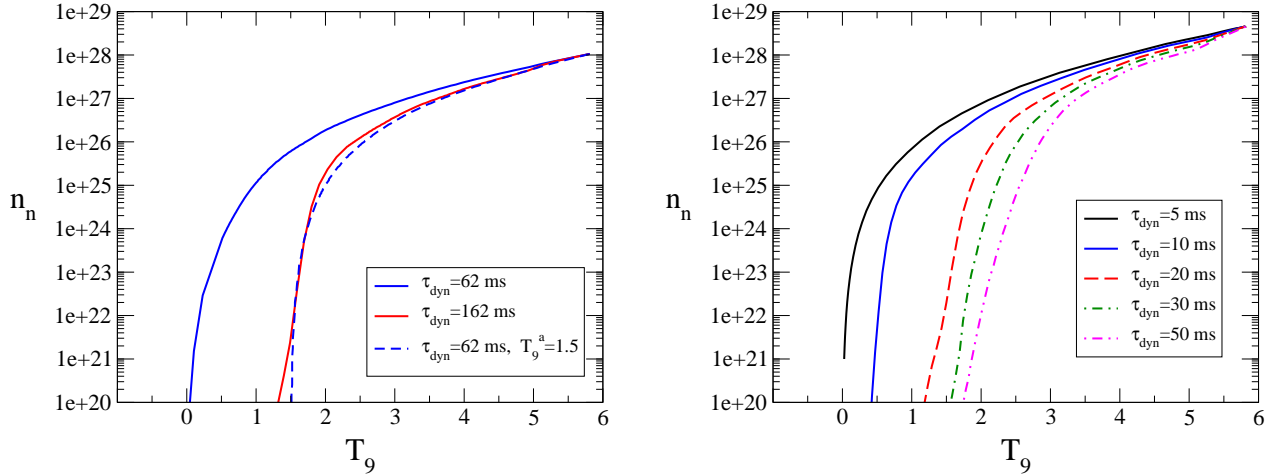

\begin{center}
\includegraphics*[width=8.0cm]{n390_4.eps} 
\hspace{0.5cm}
\includegraphics*[width=8.0cm]{ne145TNn2.eps} 
\end{center}
\caption{$T_9$-dependence of the free neutron density $n_\indis{n}$
(in particles per cm$^{-3}$) for
$s=390$ (measured in $k_{\mathrm{B}}$ per nucleon), $Y_\indis{e}=0.46$
(left) and $s=145$, $Y_\indis{e}=0.42$ (right). The different curves
correspond to different exponential expansion timescales
$\tau_\indis{dyn}$ as given in the list. The dashed line in the left
plot shows the result for conditions close to those in Fig.~7 of Witti
et al.\ (\cite{wjt94}). In the case of the HT-model considered by Witti
et al.\ (\cite{wjt94}), the temperature asymptotes to a value of
$T_9^\mathrm{a}=1.5$. The density also decreases so slowly that a high
neutron number density is supported long enough for the creation of a
strong third r-process peak. In contrast, during the free expansion of
very high entropy material with an exponential timescale of 0.62$\,$ms,
the matter dilutes too fast to produce large amounts of heavy elements,
and no prominent third r-process abundance peak can be formed. In spite
of a similar decline of the neutron density with temperature for an
expansion timescale of 162$\,$ms, the third abundance peak can also not
be assembled in this case, because the slow expansion leads to abundant
seed production in a strong $\alpha$-process and too small a
neutron-to-seed ratio.
 }
\label{fig2twj}
\end{figure*}

We compared the seed production as calculated with our full
network to results from nuclear statistical equilibrium (NSE) as obtained by
Nadyozhin \& Yudin (2004), Blinnikov et al.\ (to be published), and with the
NSE-code used by Witti et al.\ ( \cite{wjt94}). We observed rather
good agreement at temperatures $T_9 \approx 5$--6 and a density
around $5\times 10^6\,$g$\,$cm$^{-3}$.
At such temperatures the $\alpha$-peak is already formed and the
mass fraction of seed nuclei (i.e., heavy elements) is still small,
the most abundant nuclei in both cases being $^{50}$Ti and $^{54}$Cr.
Small discrepancies between the different calculations emerge
mostly from the use of different mass formulas and can lead to
smaller differences in the subsequent $\alpha$- and r-process. They
may be important for exact predictions of the abundance flow during
the r-processing, but they should not affect the basic results of our
investigation.

We also compared our results for the $\alpha$-process with the calculations
made by Witti et al.\ (\cite{wjt94}), using conditions similar to those
of the HT-model of Witti et al.\ (\cite{witti93}), i.e.,
$s=390$, $Y_\indis{e}=0.455$. For an exponential expansion with a
timescale of $\tau_{\mathrm{dyn}} = 62\,$ms, which is the same as
in the HT-model, the number density $n_\indis{n}$ as a function of
temperature (Fig.~\ref{fig2twj}, left) is clearly different from what
was obtained by Takahashi et al.\ (\cite{twj94}) (see Fig.~2 therein and the
dashed line in the left panel of Fig.~\ref{fig2twj}). When we increase
the expansion timescale to $\tau_\indis{dyn}= 162\,$ms, our result
becomes close to that of Takahashi et al.\ (\cite{twj94}). The reason
for this discrepancy is the fact that our calculations assume
homologous expansion with an exponential decrease of density and
temperature, whereas the temperature and density in the hydrodynamic
model considered by Takahashi et al.\ asymptote to nonvanishing
values, i.e., the temperature approaches $T_9 = 1.5$ during the
later expansion. Although r-process elements are formed, neither
of our exponential runs leads to the build-up of a strong third
r-process peak, despite the high entropy. While in the case with the
short expansion timescale ($\tau_{\mathrm{dyn}} = 62\,$ms)
the rapid dilution of the matter prohibits the efficient formation of seed
nuclei, the larger expansion timescale ($\tau_{\mathrm{dyn}} = 162\,$ms)
leads to a strong $\alpha$-process with a lot of seed production and
therefore a neutron-to-seed ratio that is too small for the generation
of very heavy r-process nuclei. In contrast, Takahashi et al.\
(\cite{twj94}) obtained a prominant third abundance peak. In their
calculation a sufficiently fast initial expansion prevents the
formation of too much seed material before the r-process starts,
and therefore the neutron-to-seed ratio remains high. The asymptoting
temperature and density on the other hand support a high neutron number
density for such a long time that neutron capture reactions can
assemble nuclei also in the third peak.

\begin{figure*}[tpb!]
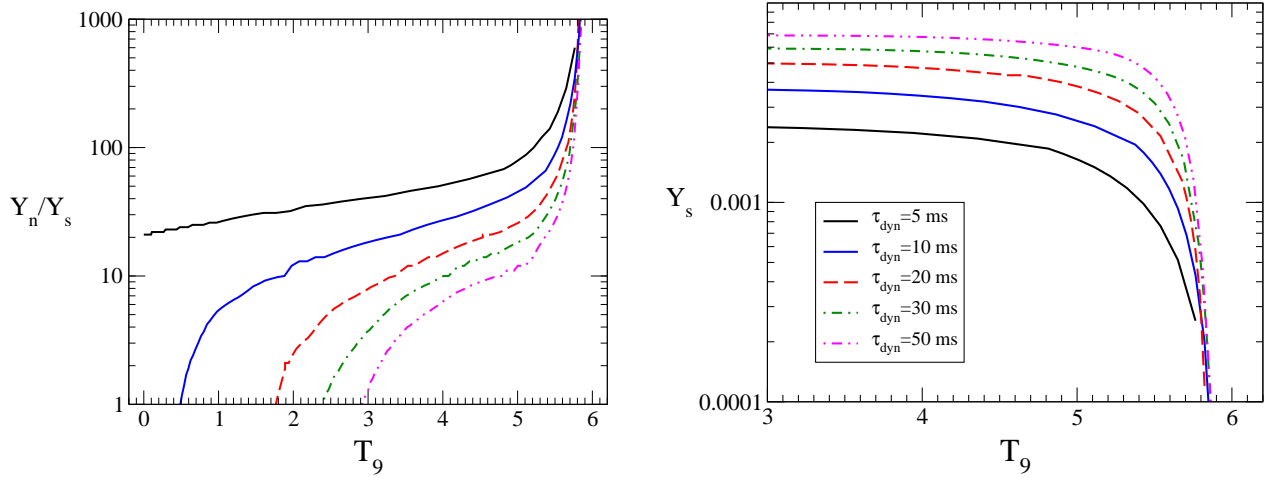

\begin{center}
\includegraphics*[width=8cm]{ne145_6_ns.eps}      
\hspace{0.5cm}
\includegraphics*[width=8cm]{ne145_6_yz.eps}    
\end{center}
\caption{$T_9$-dependence of the neutron-to-seed ratio
$Y_\indis{n}/Y_\indis{s}$ (left) and of the ``seed'' number
fraction $Y_\indis{s}$ (right) for an entropy of
$s=145$ and $Y_\indis{e}=0.42$. The different curves correspond to
different exponential expansion timescales,
$\tau_\indis{dyn}=5,10,20,30,50\,$ms (the same cases as in the
right panel of Fig.~\ref{fig2twj}).
At the beginning of r-processing
($T_9 \approx 2$), the free neutron number is sufficiently large to form
heavy elements up to the second or even third peak if the dynamical
time is less than 10$\,$ms for the considered entropy and electron fraction.
For larger values of
$\tau_\indis{dyn}$, the free neutrons are exhausted right at the beginning
of the r-processing and the number of seed nuclei is too high.
}
\label{ynys_t9}
\end{figure*}

For lower entropy, $s=145$, and higher neutron excess,
$Y_\indis{e}=0.42$, (Fig.~\ref{fig2twj}, right) the number of free
neutrons can also remain large enough to allow for the onset of
r-processing when the $\alpha$-process freezes out at a temperature of
$T_9\sim 2$. In this case, however,
much shorter dynamical timescales are needed. Figure~\ref{ynys_t9}
displays the decreasing seed formation and increasing neutron-to-seed
ratio for smaller expansion timescales $\tau_{\mathrm{dyn}}$.

%
%

How many neutrons per seed nucleus are needed to form the third abundance
peak ($A\sim 196$)? Usually the number mentioned in this context is
not less than 150. But this estimate is based on the simple calculation
how many neutrons a single $^{56}$Fe nucleus must capture to finally,
after a chain of beta-decays, end as nucleus in the platinum peak.
In the actual r-process, only a fraction of the nuclei that are
initially formed by the alpha-process --- we call these nuclei the ``seed''
--- will be ultimately transformed to elements and isotopes
with the highest mass numbers.

Taking into account that the observed ratio of the third to second
abundance peak is $Y_{196}/Y_{130}\sim 0.2$, we estimate that the
neutron number needed to obtain a strong third peak is of the order of
$(196-60)\times 0.2\approx 30$. The actual value might even be a bit
lower, because the mean atomic number at the freeze-out time of the
$\alpha$-process can be about 80 or more instead of 60 (an exact
estimate, however, also depends on the conditions in which the
r-processing takes place and the corresponding speed of the abundance
flow). Therefore a solar-like formation of the platinum peak can be
expected if one has a neutron-to-seed ratio of around 30 {\em after the
freeze-out of charged-particle reactions}.

We point out, however, that this value does not ensure that the
solar abundance distribution in the whole mass-number region between 
the second and third abundance peaks is reproduced well. This goal is
not attempted in our study here, and it is well known that
a superposition of at least two components with different physical 
conditions instead of a single ejecta trajectory (as considered by us)
is needed for matching the solar r-process pattern (see, e.g.,
Kratz et al.\ 1993, Goriely \& Arnould 1996, Wanajo et al.\ 2004, 
Kuroda et al.\ 2008). In our fast-expansion cases we regard the
described criterion as a reasonably good indicator for outflow
conditions that allow for the appearance of a strong third peak
during the r-processing that follows after the freeze-out of 
charged-particle reactions.

To judge about the possibility of an r-process for different choices
of the parameters of our dynamical model ($s$, $Y_\mathrm{e}$,
$\tau_\mathrm{dyn}$), we first consider the seed formation and the
corresponding time evolution of the neutron-to-seed ratio in
the homologous expansion phase. In Fig.~\ref{fig2twj} we have seen
agreement of the free neutron density as a function of time between
our calculations and the prior ones by Takahashi et al.\ (1994) for
a suitable choice of the expansion timescale $\tau_\mathrm{dyn}$
(although the heavy-element nucleosynthesis was considerably
different as discussed above).
Figure~\ref{ynys_t9} shows the neutron-to-seed ratios and the seed
abundances, $Y_\indis{s}=\sum_{Z>2}Y_{Z}$, versus
temperature for the same cases as displayed in the
right panel of Fig.~\ref{fig2twj}. Note that we consider all nuclei
with $Z>2$ as seeds, which is a different definition than used by
Terasawa et al.\ (\cite{tera01}), who restriced seed nuclei to the
more narrow range of $70\leq A\leq 120$ and $Z>26$. The latter mass range
is appropriate when the formation of the second abundance peak in an
incomplete r-process is discussed. Here, however, we explore the
possibility of third-peak production and consider a combined
$\alpha$- and r-process. In this case the atomic mass number range
should not be constrained. Because of the very short expansion timescale
and the treatment of both the $\alpha$-process and the r-process
within the framework of the same mathematical model and computer code,
a distinction of seed nuclei and r-processed material becomes
artificial and one has to carefully judge when this information is
measured and what it means for the evolution of the mass number
with time. Often it refers to older calculations in which
the $\alpha$- and r-process were computed in two independent steps
and with different codes. In this context we note that the amount
of heavy nuclei in our calculations is approximately twice as big
as found by Terasawa et al.\ (2002). This difference is mostly explained
by the different model parameters (in particular, Terasawa et al.
investigated outflows with higher entropies), but to some extent may
also be a consequence of different nuclear rates and included reactions.

With the definition of seed nuclei adopted by us, the seed abundance
becomes constant after the freeze-out of the alpha-process and will
also be unaffected by the transition to a second stage with modified
expansion behavior for the cases considered by us. In contrast,
replacing the free (homologous) expansion by a slower second stage of
outflow behavior will have an impact on the density of free neutrons
as a function of time. The free neutron density determines the
r-process path and thus the formation of the heaviest elements in
the third abundance peak.

 From Fig.~\ref{ynys_t9} (left panel) we see that the expansion
timescale for third peak formation 
to happen should be less than about 10$\,$ms. In
this case the neutron-to seed ratio $Y_\indis{n}/Y_\indis{s}$ as the
decisive macroscopic factor for the platinum peak formation reaches the
interesting values mentioned above. Naturally, this ratio depends not
only on the number of free neutrons, but also on the seed abundance.
Figure~\ref{ynys_t9} (right panel) shows that the seed production
varies strongly with the expansion timescale: if the dynamical time is
large, a lot of seed is assembled and all neutrons will be captured
faster than the platinum peak begins to appear. With shorter expansion
timescale the seed production drops and at the same time the density of
free neutrons
increases (see right panel of Fig.~\ref{fig2twj}). The
consequence of both trends with reduced $\tau_{\mathrm{dyn}}$ is a
strong growth of the $Y_\indis{n}/Y_\indis{s}$ ratio that is present at
the beginning of the r-process.

However, as already discussed above in the context of
Fig.~\ref{fig2twj}, too rapid expansion can also be disadvantageous for
a strong r-process. When the dynamical timescale becomes very short,
the expansion and gas dilution proceed faster than the rate of
recombination of alpha particles to seed nuclei and subsequently the
rate of neutron captures. For the conditions considered in
Fig.~\ref{ynys_t9} this happens   when the homologous expansion and
density decrease are too rapid to allow for the neutrons to be
assembled into a strong 3rd peak, despite a high neutron-to-seed ratio
and a sufficiently small seed concentration $Y_\indis{s}$ below a
critical upper value of $Y_\indis{s}^\mathrm{cr} \sim 3\times 10^{-3}$
after the freeze-out of charged-particle processes.

Of course, as discussed in detail in many previous works, besides
the expansion timescale, the entropy and the neutron excess (or
$Y_\mathrm{e}$), have a sensitive influence on the strength of
the r-processing, i.e., on the question how many nuclei with mass
numbers above the second abundance peak and in particular near the
third peak can be formed. In any case, for rapid expansion (small
$\tau_\mathrm{dyn}$) the seed production is reduced and the
neutron-to-seed ratio becomes more favorable for a strong r-process.
The heavier the seed nuclei at the end of the $\alpha$-process are,
the lower can the required number of free neutrons be. This was
already discussed by, e.g., Panov \& Chechetkin (2002), who showed
that fairly low neutron/seed ratios are already sufficient when the
seed material at the freeze-out time of charged-particle reactions
consists mostly of nuclei in the second abundance peak.

\begin{figure}[tpb!]
 \begin{center}
 \includegraphics*[width=8cm]{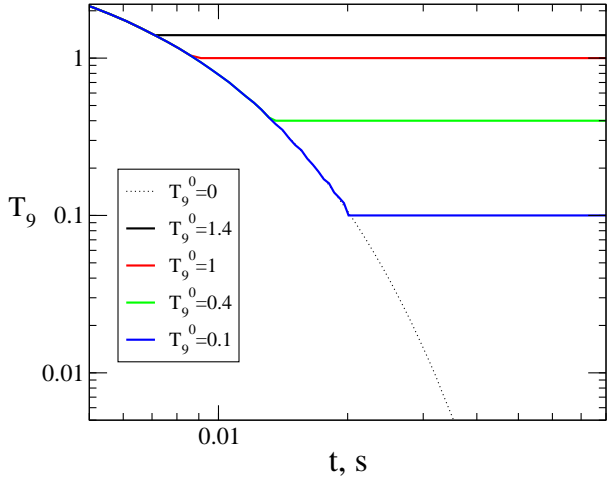}
 \end{center}
  \caption{Time evolution of the temperature for the different
  considered cases with exponential first expansion phase
  (starting at $T_9 \sim 6$ and declining with a dynamical
  timescale of $\tau_{\mathrm{dyn}} =5\,$ms)
  and constant later phase.}
  \label{nn_t}
\end{figure}

\section{Asymptotic behavior of temperature and density and
formation of the platinum peak }

Our combined ($\alpha$+r)-code was applied to
nucleosynthesis calculations assuming the two-stage expansion
behavior described above, with a second phase of either constant
or slowly decreasing temperature and density following a first
phase of rapid, exponential expansion. Our calculations were
started at a temperature of $T_9=$~6, assuming NSE at this point.
For conditions similar to those given in Fig.~2
of Terasawa et al.\ (2001) with a short expansion timescale
of $\tau_\mathrm{dyn} = 5\,$ms,
we obtained basically the same results for the development of
the neutron number density and average neutron separation energy
until about half a second, and
found gross agreement of the structure of the
abundance distribution. Smaller discrepancies might be attributed
to differences in the initial composition and nuclear reaction
rates. The number fraction of seed nuclei reported by
Terasawa et al.\ (2001), $Y_\mathrm{s}\sim 0.001$, is a bit
low compared to our results, but this is probably mainly caused
by a different definition of ``seeds''.

In contrast, we were not able to confirm the formation of the
third abundance peak for conditions similar to those considered
by Terasawa et al.\ (2002) with an assumed expansion timescale of
$\tau_\mathrm{dyn} = 25\,$ms. Besides not providing exact
information about the initial density and the kind of network
used for the alpha- and r-process calculations (a full network
or off-line calculations for the r-processing?), some of the
results and explanations are hard to reproduce in detail.
For example, the calculations done by us with different NSE
codes (including that of Nadyozhin and Yudin 2004) and also
the simulations by Witti et al.\ (1994) show that
$\alpha$-particles reassemble mainly down to temperatures around
$T_9 = 5$--6, somewhat dependent on the density, but not at
temperatures as low as $T_9 = 4$ (see Fig.~1 in Terasawa et al.\
2002). Moreover, they argued that a lower value of the asymptotic
temperature is favorable for a successful r-processing by allowing
for a higher neutron-to-seed ratio, because such a lower ``outer
boundary'' temperature reduces charged-particle
reactions and thus the production of seed nuclei. This suggested
influence of the chosen asymptotic temperatures in the range between
$T_9 = 0.4$ and $T_9 = 1.3$, however, is implausible, because
charged-particle reactions become inefficient already at temperatures
of $T_9 \approx 2$, when the thermal energies of protons and
$\alpha$-particles become too low for enabling these particles
to overcome the nuclear Coulomb barriers.

We will therefore not further attempt to compare
our calculations with those done by Terasawa et al.\ (2002). Instead,
we will in the following present our results for the formation of
the third abundance peak in dependence of entropy, electron fraction,
and dynamical timescale during the first, exponential expansion phase.
In particular we will also study variations with the asymptotic values
of temperature and density during the second, slower phase of expansion.
This will help us in the analysis of the physical effects that can
explain our results. Our goal is to develop a deeper understanding of
the influence of the late-time behavior of neutrino-driven winds on
the possibility of strong r-processing in such an environment.

To this end we carried out a set of calculations for exponential
timescales in the range from 1.0 to 25$\,$ms and for four values of
the wind entropy (in units of Boltzmann's constant per nucleon),
$s=105$, 145, 170, and 200. The initial neutron to
proton ratio was determined by an electron fraction of usually
$Y_\indis{e}=0.42$; some runs were performed with a value of
$Y_\indis{e}=0.46$.

Our special attention was on the second stage of slower expansion,
which was chosen to either proceed with constant temperature and
density (Eqs.~\ref{eq:late0}--\ref{eq:late1r}) or gradually decreasing
temperature and density (Eqs.~\ref{eq:late2}--\ref{eq:late3r}).
In the former case we defined the asymptotic
temperature as $T_9^\mathrm{f}(t \geq t_0)\equiv T_9^0 =
\mathrm{const}$ (Sect.~\ref{sec:constT}),
in the latter case the temperature was assumed to follow a
power-law time-dependence according to
$T_9^\mathrm{f}(t)\equiv T_9^\mathrm{f}(t_0)\times (t_0/t)^{2/3}$ for
$t \geq t_0$ (Sect.~\ref{sec:power-law}).
The range of temperatures $T_9^\mathrm{f}(t_0)$, where
the second expansion phase began, had a broad overlap with
the ``boundary temperatures'' considered by Terasawa et al.\ (2001,
2002) and Wanajo et al.\ (2002).

Our standard calculations were performed with an entropy of $s = 145$.
Values of less than $s\approx 100$ turned out to lead only to the
production of the second abundance
peak around $A\sim 130$, but r-processing up to the platinum
peak was not possible when the other characteristic wind parameters
were varied within the limits mentioned above.

\begin{figure*}[tpb!]
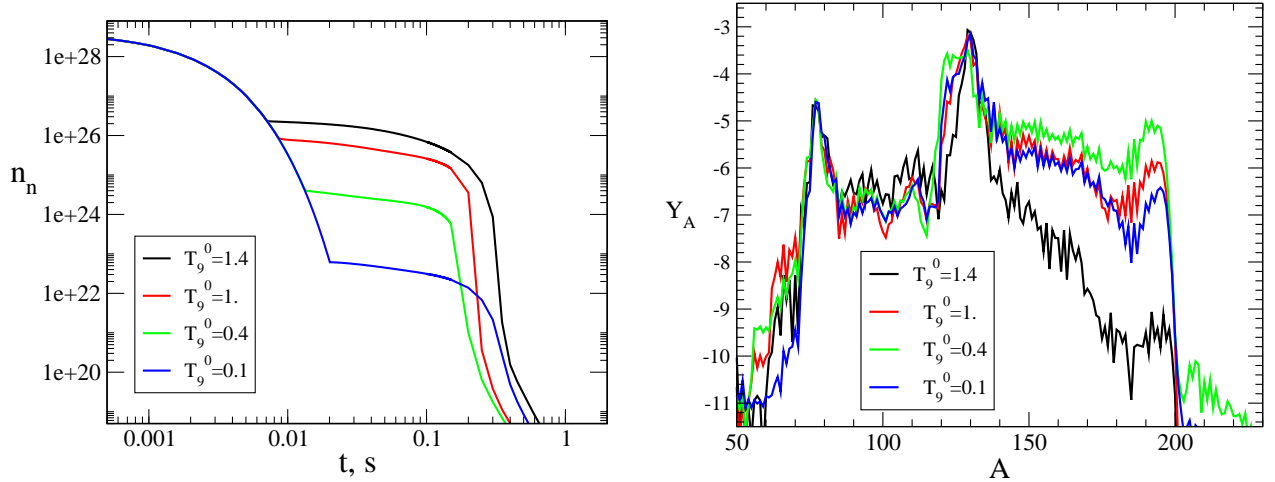

 \begin{center}
 \includegraphics*[width=8cm]{nc145t5nn.eps}    
 \hspace{0.5cm}
 \includegraphics*[width=8cm]{nc145t5ya.eps}    
 \end{center}
  \caption{Time dependence of the neutron number density,
  $n_\indis{n}(t)$ (left, in particles per cm$^{-3}$),
  and the final abundance distributions,
  $Y_\indis{A}$, resulting from our network calculations for
  $s=145$, $Y_\indis{e}=0.42$, a short dynamical timescale of
  $\tau_{\mathrm{dyn}} =5\,$ms, and different values of the
  constant asymptotic temperature $T_{9}^\mathrm{f}(t\ge t_0)=T_9^0(t_0)$
  during the second stage of the ejecta expansion.
  The corresponding values of $T_9^0$ are listed in the inset and
  the temperature evolution of function of time is shown in 
  Fig.~\ref{nn_t}.}
  \label{nn_t145}
\end{figure*}

\subsection{Constant asymptotic temperature and density}
\label{sec:constT}

\begin{figure*}[tpb!]
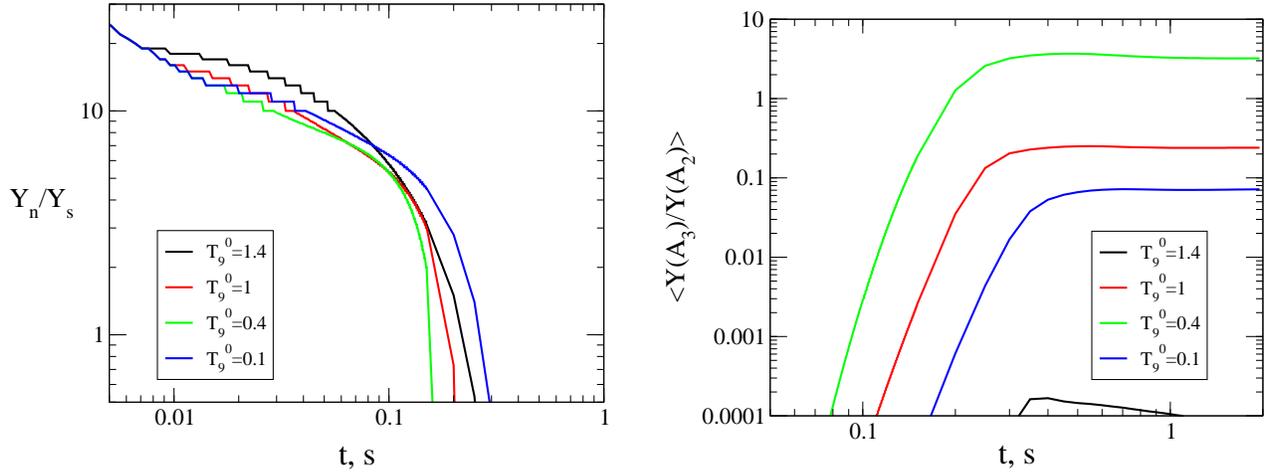

\begin{center}
 \includegraphics*[width=8cm]{fc145t5nS.eps}
\hspace{0.5cm}
 \includegraphics*[width=8cm]{fc145t5_A3.eps}
\end{center}
\caption{Time dependence of the neutron-to-seed ratios,
$Y_\indis{n}/Y_\indis{s}$
({\em left}), and of the height of the third abundance peak relative to the
second, $\left\langle Y(A_3)/Y(A_2)\right\rangle$ ({\em right}, in percent)
for the same conditions as in Fig.~\ref{nn_t145}
($s=145$, $Y_\indis{e}=0.42$, $\tau_\indis{dyn}=5$~ms,
$T_9^0={\mathrm{const}}$ as given in the inset).
The inversion of the third peak
formation with decreasing asymptotic temperature can be clearly seen:
when $T_9^0$ is lowered from 1.4 to 0.4,
the ratio of the third peak to the second increases, and when $T_9^0$
is reduced even further, $\left\langle Y(A_3)/Y(A_2)\right\rangle$ begins to
drop again.}
  \label{ynys_tc}
\end{figure*}

\begin{figure*}[tpb!]
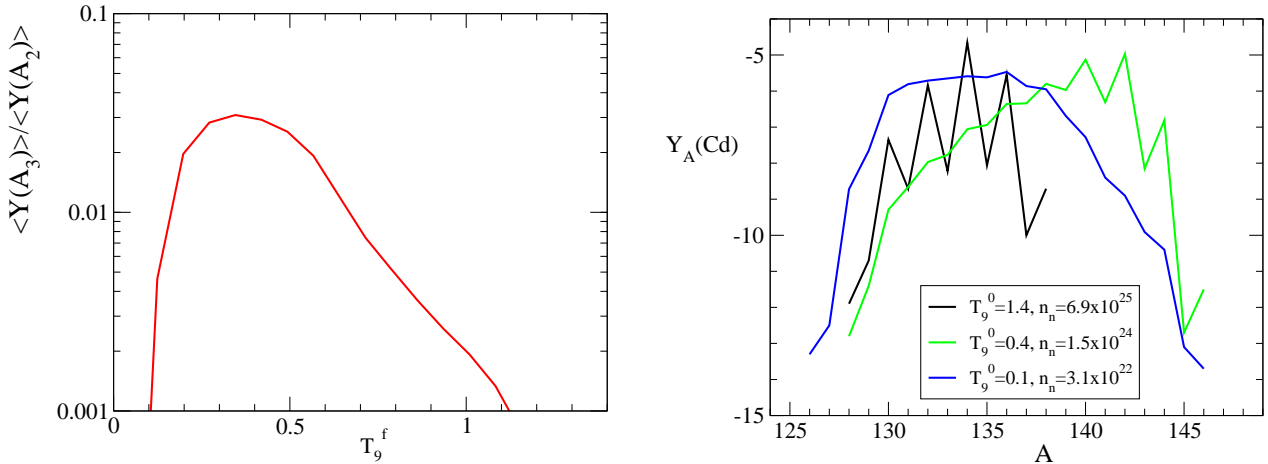

\begin{center}
\includegraphics*[width=8cm]{prof_a3f2.eps}
\hspace{0.5cm}
\includegraphics*[width=8cm]{cd_ya_a2n.eps}
\end{center}
\caption{{\em Left:} The changing strength of the third abundance peak,
measured in terms of the ratio $\left\langle
Y(A_3)/Y(A_2)\right\rangle$, as function of the asymptotic temperature
$T_9^\mathrm{f}(t > t_0) = T_9^0$ for the conditions considered in
Fig.~\ref{nn_t145} ($s=145$, $Y_\indis{e}=0.42$,
$\tau_\indis{dyn}=5$~ms, $T_9^0={\mathrm{const}}$). The relative height
of the third abundance peak depends strongly on the value of $T_9^0$
because of a sensitive influence of this temperature on the r-process
path. This can be seen in the distribution of cadmium isotopes at an
evolution time of $t = 0.1\,$s for three different cases of the
asymptotic temperature ({\em right}; the neutron number densities in
the inset are those at $t = 0.1\,$s in Fig.~\ref{nn_t145}). When
$T_9^0$ is lowered from 1.4 to 0.4, the free neutron density and thus
the neutron capture rate in the slow, second phase of the expansion
decreases less strongly than the ($\gamma$,n)-rates, which are
extremely sensitive to the temperature. As a consequence, the r-process
path is shifted towards the neutron-drip line (compare the isotope
distribution given by the green line) where the $\beta$-decay rates are
higher and the r-processing proceeds so rapidly that there is time to
form a strong third abundance peak. For even lower asymptotic
temperature the ($\gamma$,n)-rates do not play an important role and
the strength of the r-processing is determined by the free neutron
density and the $\beta$-decay rates. Since the r-process path returns
again to a location closer to the valley of stability (see the isotope
distribution of the blue line), the $\beta$-decay rates are lower and
the nucleosynthesis slows down. Therefore a strong third peak cannot be
formed. } \label{sn_t9_nn}
\end{figure*}

In this section we consider an exponential first expansion phase
that is superseded at evolution time $t_0$ by a second, slow
expansion phase whose asymptotic velocity at large radii goes to
zero. In this case the density and temperature during the second
phase adopt constant values (Eqs.~\ref{eq:late0}--\ref{eq:late1r}).
We explore various choices of the asymptotic temperature
$T_9^\mathrm{f}(t \geq t_0) \equiv T_9^0 = \mathrm{const}$ between
0.1 and 1.4 (see Fig.~\ref{nn_t}). 
For an entropy of $s= 145$, for which results are
displayed in Figs.~\ref{nn_t145}--\ref{sn_t9_nn}, these temperatures
correspond to constant densities $\rho_0$ between about
1$\,$g$\,$cm$^{-3}$ and $10^4\,$g$\,$cm$^{-3}$. For these values the
neutron number density at time $t_0$ shows differences by more than
three orders of magnitude (see Fig.~\ref{nn_t145}). Although the
lowest assumed value of $T_9=0.1$ might appear extreme, because it
requires very rapid exponential expansion for a longer time with
fairly high velocities at the end of this phase
  (see Table~\ref{tab:parameters})  ,
the wide range of asymptotic temperatures allows us to
better understand the dependence of the nucleosynthesis on the
late-time expansion behavior of the outflowing matter.

When the asymptotic temperature is as high as $T_9^0 \geq 1.4$, the
third peak hardly develops (see the abundance distribution at the end
of our calculations in the right panel of Fig.~\ref{nn_t145}). Only
when the asymptotic temperature is reduced from this value to smaller
numbers, the platinum peak grows in strength, because the free neutrons
are exhausted more rapidly and the r-processing proceeds faster towards
the high mass-number region. The latter fact can be seen by comparing
the decrease of the neutron density with time for $T_9^0 = 1.4$, 1, and
0.4 in the left panel of Fig.~\ref{nn_t145}, and it is also visible
from the neutron-to-seed ratios $Y_\mathrm{n}/Y_{\mathrm{s}}$ as
functions of time in the left panel of Fig.~\ref{ynys_tc}. When the
asymptotic temperature is lowered to less than $T_9^0 \approx 0.3$, the
free neutron density $n_\mathrm{n}$ during the slow-expansion phase
becomes so low that the formation of the third abundance peak under
such conditions slows down considerably. Correspondingly,
$n_\mathrm{n}$ as well as $Y_\mathrm{n}/Y_{\mathrm{s}}$ decrease less
quickly (see again the left panels of Figs.~\ref{nn_t145} and
\ref{ynys_tc}) and the height of the third peak at the time of neutron
exhaustion becomes clearly lower (Fig.~\ref{nn_t145}, right panel).

This inversion of the third peak formation with decreasing
value of the asymptotic temperature $T_9^0$
can be clearly seen from the
time evolution of the average of the yields in the platinum peak
relative to those in the $N=82$ peak,
$\left\langle Y(A_3)/Y(A_2)\right\rangle$,
in the right panel of Fig.~\ref{ynys_tc}.
The displayed quantity is defined by summing up the produced yields
for five mass numbers around $A=130$ and $A=196$ and then computing
the ratio\footnote{A similar peak ratio was considered in a
recent paper by Beun et al.\ (2008).}
\begin{equation}
\left\langle {Y(A_3)\over Y(A_2)}\right\rangle\, \equiv\,
{\sum_{193}^{197}Y(A_i)\over \sum_{128}^{132}Y(A_i)}\, .
\label{eq:peak_heights}
\end{equation}
The left panel of
Fig.~\ref{sn_t9_nn} shows the sensitivity of the final value of this
ratio to the chosen temperature $T_9^\mathrm{f}(t > t_0) = T_9^0$.
While the growing strength of the platinum peak with lower asymptotic
temperature agrees with the finding by Terasawa et al.\ (2002), their
conclusion of a more efficient r-processing for lower ``outer boundary
pressure/temperature'' obviously does not hold any more when this
temperature is less than $T_9 \approx 0.3$, which is below the range
explored by them.

We note here that often the change of the average atomic number
is used for following the assembling of heavy elements beyond the
$A\approx 130$ abundance peak. We found this quantity to be less
suitable for this purpose than our height ratio
$\left\langle Y(A_3)/Y(A_2)\right\rangle$, in particular when
the second peak is much stronger than the third. A growth of
the latter by one order of magnitude can mean a change of
$\left\langle A\right\rangle$ by just one or a few units,
whereas $\left\langle Y(A_3)/Y(A_2)\right\rangle$ varies sensitively
and thus serves well as a tracer of changes of the abundance
distribution in the high-mass number region.
The numbers listed in Table~\ref{tab:parameters}, for example,
confirm the usefulness, in fact the superiority, of the ratio 
$\left\langle Y(A_3)/Y(A_2)\right\rangle$ instead of the mean mass
number, $\left\langle A \right\rangle$, as an indicator of the third
peak formation at least in the cases with very fast outflow expansion
considered here (of course, again we do not claim that a single parameter
value or a single outflow trajectory is sufficient for getting a good match
of the whole r-process abundance distribution between the second and
third peaks). While $\left\langle Y(A_3)/Y(A_2)\right\rangle$ exhibits 
a rapid growth during the build-up of the third abundance peak and
changes by many orders of magnitude between cases with weak or strong
third peak, the average mass number remains within the relatively 
narrow interval of $118 \le \left\langle A\right\rangle\le 145$
(see Table~\ref{tab:parameters}).

The variation of the strength of the third peak with different
asymptotic temperatures can be understood from the sensitivity of the
neutron capture rates and nuclear photodisintegration rates to the
neutron number density and temperature, respectively, and by the
competition of these rates. This competition determines the location of
the r-process path and thus the speed of the nucleosynthesis, which is
defined by the $\beta$-decay rates. When the temperature during the
second, slow expansion phase is large ($T_9^0 \ga 1.0$), the
($\gamma$,n)-reactions are very fast and the r-process path   lies
closer to the stability region than in case of smaller $T_9^0$.    The
r-process flow beyond the second peak is then rather weak because of
the low beta-decay rates and a correspondingly slow progression of the
nuclear flow. When the asymptotic temperature is reduced to $T_9^0
\approx 0.2$--0.7, the ($\gamma$,n)-rates decrease and the r-process
path moves towards the neutron-drip line where the $\beta$-decay rates
are higher. Therefore the r-processing proceeds faster beyond the
second peak, leading to a more rapid drop of the free neutron density
and a more efficient third peak production. When the asymptotic
temperature is lowered to less than $T_9^0 \approx 0.2$ during the late
expansion phase, the neutron densities are very low so that the
r-process path returns to a location closer to the $\beta$-stable
region. In this case the $\beta$-decays again become slower and
therefore the r-process nucleosynthesis decelerates and the third peak
builds up to a smaller height.

\begin{figure}[tpb!]
 \begin{center}
 \includegraphics*[width=8cm]{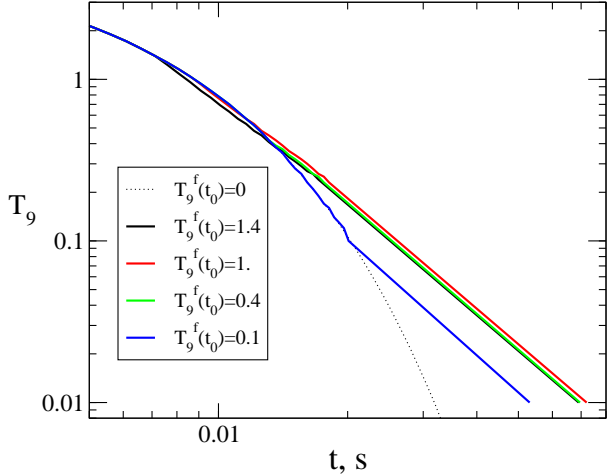}
 \end{center}
  \caption{Time evolution of the temperature for the different
  considered cases with exponential first expansion phase
  (starting at $T_9 \sim 6$ and declining with a dynamical
  timescale of $\tau_{\mathrm{dyn}} =5\,$ms)
  and power-law later phase.}
  \label{nn_t2}
\end{figure}

These movements of the r-process path are a consequence of the
different influence of a change of the asymptotic temperature
on the ($\gamma$,n)- and the neutron captures rates.
For asymptotic temperatures in the interval
$0.2\la T_9^0\la 1.0$ the ($\gamma$,n)-rates are very sensitive
to temperature variations. In contrast, the neutron capture rates change
with the corresponding variations of the density and free neutron
density less strongly. A reduced temperature therefore decreases
the photodisintegration rates significantly, whereas the neutron
captures remain fast despite the lower neutron density. When
$T_9^0\la 0.2$, the ($\gamma$,n)-rates become less relevant,
but neutron captures still compete with $\beta$-decays. For such
low temperatures and thus low neutron densities, the neutron
capture rates are too small to drive the r-process path far away
from the valley of stability.

The shift of the r-process path that is caused by different
asymptotic temperatures (and connected parameters) manifests
itself in different isotopic profiles of the elements formed by
the r-processing (i.e.\ in different yield of the isotopes of
an element). This can be seen for the case of cadmium in the
right panel of Fig.~\ref{sn_t9_nn}.

\subsection{Power-law time-dependence of the asymptotic temperature and density}
\label{sec:power-law}

In this section we consider the case that the
initial exponential phase is superseded at time $t = t_0$ by a
slow expansion phase in which the temperature and density decay
according to power laws, i.e.,
$T_9^\mathrm{f}(t)\propto T_9^\mathrm{f}(t_0)/t^{2/3}$ and
$\rho^\mathrm{f}(t)\propto \rho^\mathrm{f}(t_0)/t^2$
for $t \geq t_0$ (Eqs.~\ref{eq:late2}--\ref{eq:late3r}).
The temperature evolution as function of time for the considered
cases is displayed in Fig.~\ref{nn_t2}.

\begin{figure*}[tpb!]
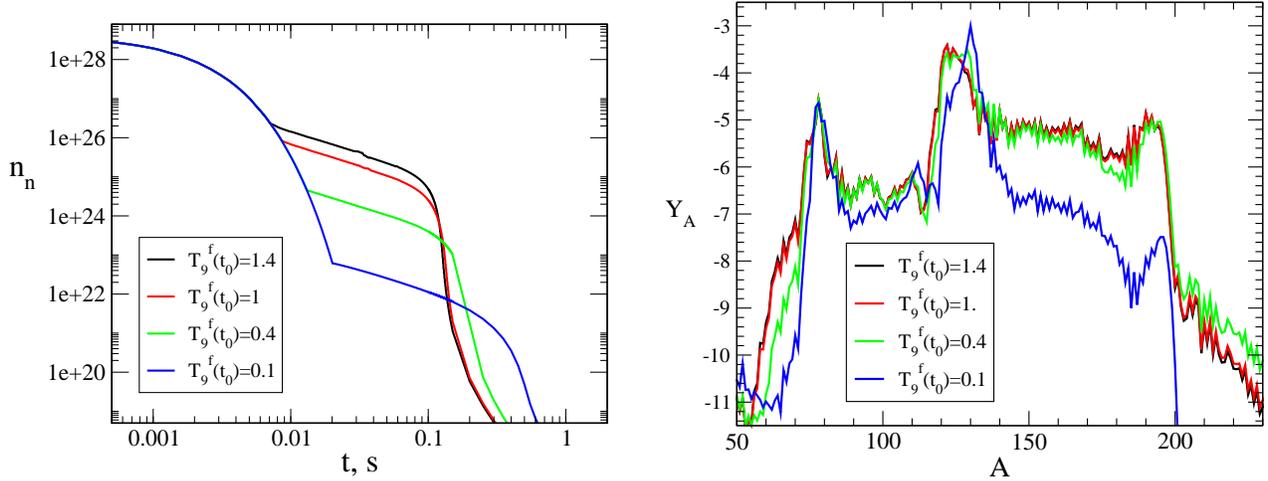

 \begin{center}
 \includegraphics*[width=8cm]{nf145t5nn.eps}
 \hspace{0.5cm}
 \includegraphics*[width=8cm]{nf145t5ya.eps}
 \end{center}
\caption{Same as Fig.~\ref{nn_t145} (with $s = 145$, $Y_\mathrm{e} = 0.42$,
and an exponential timescale $\tau_\mathrm{dyn} = 5\,$ms),
but for a power-law decay of
the temperature during the second, slow expansion phase: $T_9^\mathrm{f}(t)
\propto T_9^\mathrm{f}(t_0)/t^{2/3}$ for $t \geq t_0$ (see Fig.~\ref{nn_t2}).
The left plot shows
the time dependence of the neutron number density, $n_\indis{n}(t)$
(in particles per cm$^{-3}$),
the right plot the corresponding final abundance distribution for different
values of the transition temperature $T_9^\mathrm{f}(t_0)$ as given in
the inset. Different from the case with constant asymptotic temperature
in Fig.~\ref{nn_t145}, the temperature and thus the
photodisintegration rates drop rapidly also in the slow expansion phase.
Neutron captures
are therefore able to produce a prominent third abundance peak even for the
largest considered value of the transition temperature,
$T_9^\mathrm{f}(t_0) = 1.4$.
The height of the third peak relative to the second
is very similar in the whole range of temperatures $T_9^\mathrm{f}(t_0)$
between 0.4 and 1.4, and the inversion behavior of
$\left\langle Y(A_3)/Y(A_2)\right\rangle$ with $T_9^\mathrm{f}(t_0)$
is absent.
}
\label{ya_t145}
\end{figure*}

For our standard set of wind parameters, $s = 145$ (in units of
Boltzmann's constant per nucleon), $Y_\mathrm{e} = 0.42$,
and $\tau_\mathrm{dyn} = 5\,$ms already used in Sect.~\ref{sec:constT},
Fig.~\ref{ya_t145} shows the time evolution of the free neutron density
(left panel) and the final abundance distributions for four different
values of the transition temperature $T_9^\mathrm{f}(t_0)$ (right panel).
The formation of the third abundance peak turns out to be fairly
insensitive to variations of $T_9^\mathrm{f}(t_0)$ between about 0.4
and 1.4. For all transition temperatures in this interval 
the temperature
evolution is very similar (Fig.~\ref{nn_t2}) and a prominent
third peak appears.

Since the temperature in the post-exponential phase drops rapidly,
the strongly temperature-dependent ($\gamma$,n)-reactions become
unimportant very soon, while the considered entropy allows
a high free neutron density ($n_\indis{n} \ge 10^{22}\,$cm$^{-3}$)
to be present still for a long time. At such conditions the
r-process path moves very close to the neutron-drip line and
returns to the classical r-process path during free neutron
exhaustion (Panov 2003). For the considered conditions its location
shifts significantly only when the free neutron density changes
by 2--3 orders of magnitude. This explains the relative robustness
of the abundance yields to variations of $T_9^\mathrm{f}(t_0)$
around 1.0.

\begin{figure*}[tpb!]
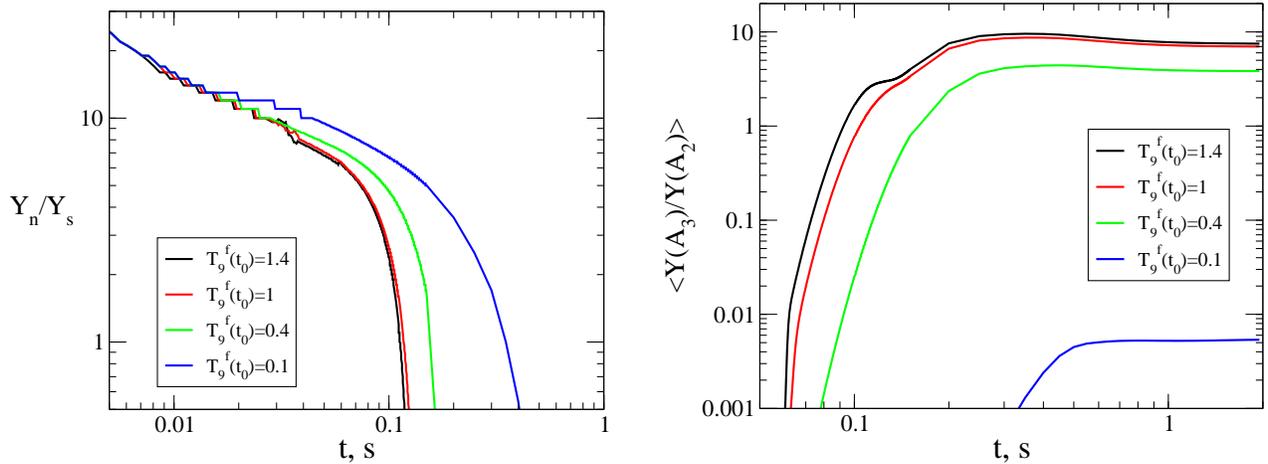

 \begin{center}
 \includegraphics*[width=8cm]{ff145t5nS.eps}
 \hspace{0.5cm}
 \includegraphics*[width=8cm]{ff145t5_A3.eps}
 \end{center}
\caption{Same as Fig.~\ref{ynys_tc} (with $s = 145$, $Y_\mathrm{e} = 0.42$,
and an exponential timescale $\tau_\mathrm{dyn} = 5\,$ms),
but for a power-law decay of
the temperature during the second, slow expansion phase: $T_9^\mathrm{f}(t)
\propto T_9^\mathrm{f}(t_0)/t^{2/3}$ for $t \geq t_0$ (as 
considered in Fig.~\ref{ya_t145}). The left plot shows
the time-dependent neutron-to-seed ratio $Y_\indis{n}/Y_\indis{s}$, the
right plot the height of the third abundance peak relative to the second
($\left\langle Y(A_3)/Y(A_2)\right\rangle$, in percent) as functions of time.
The weak dependence of the r-processing
on variations of the transition temperature $T_9^\mathrm{f}(t_0)$ between
0.4 and 1.4 as visible in the right panel of Fig.~\ref{ya_t145} is also
evident from these two plots.}
  \label{ynys_tf}
\end{figure*}

\begin{figure*}[tpb!]
 \begin{center}
 \includegraphics*[width=8cm]{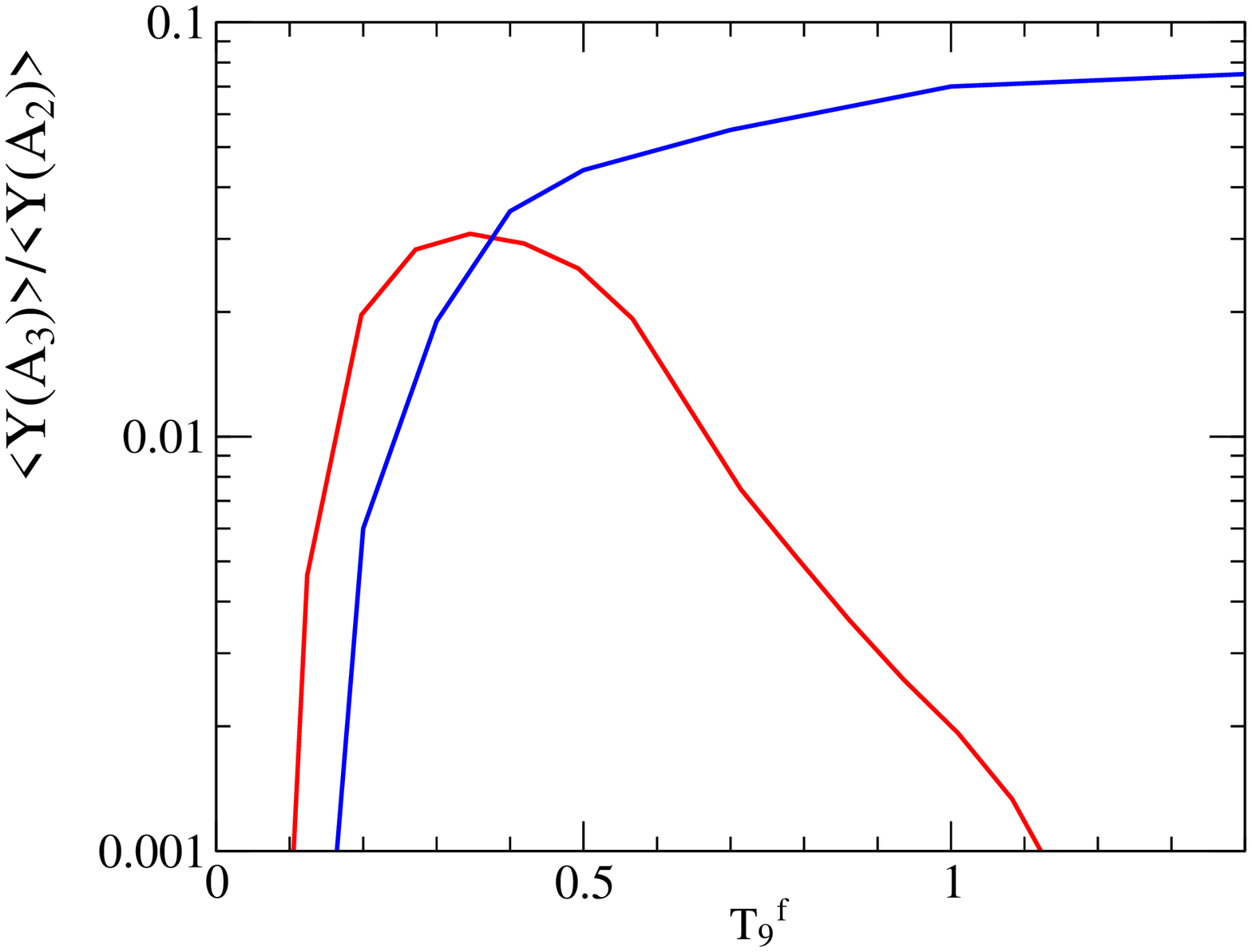}
 \hspace{0.5cm}
 \includegraphics*[width=8cm]{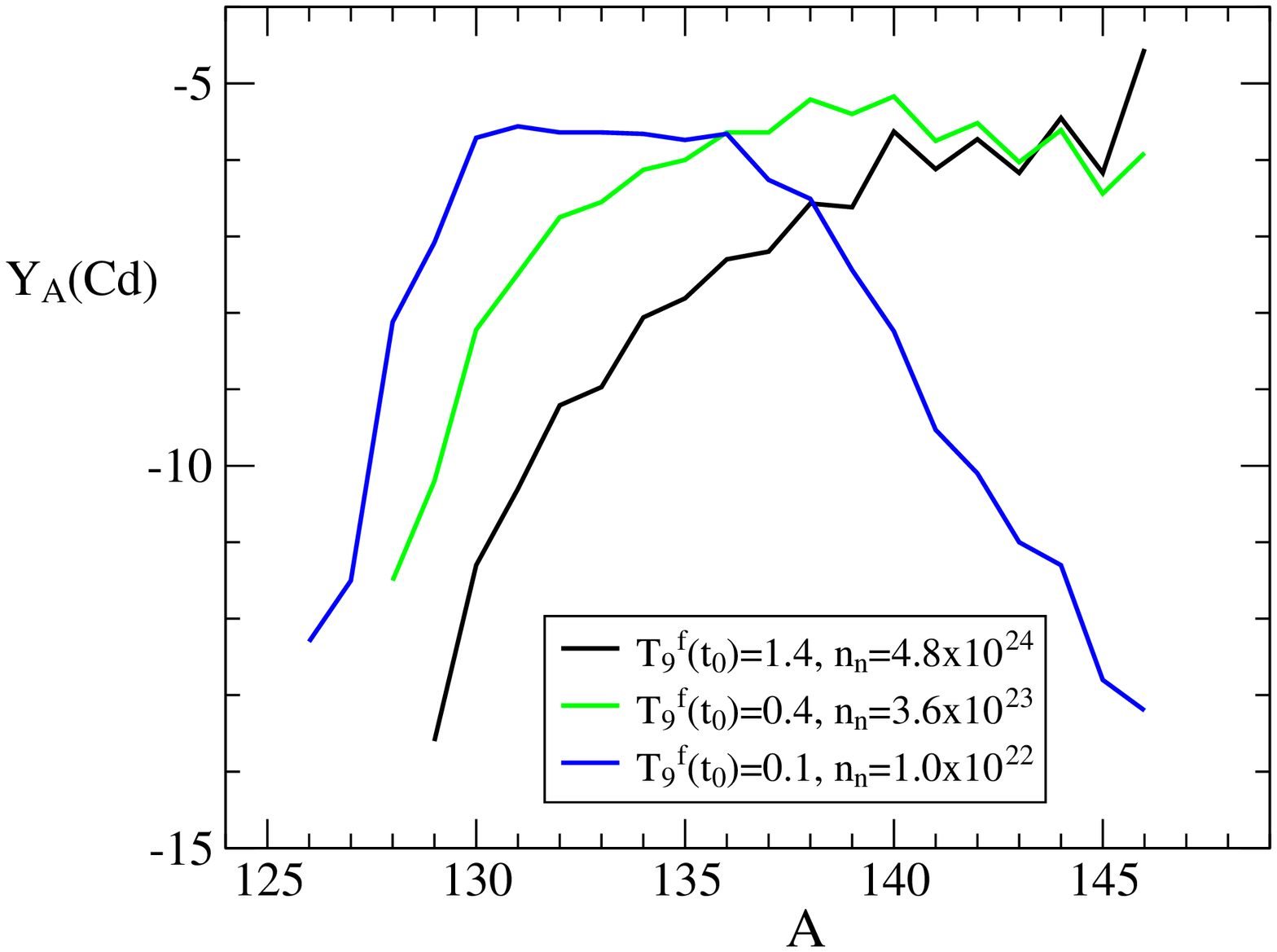}
 \end{center}
\caption{{\em Left:} Height of the third abundance peak relative to the
second,
measured in terms of the ratio $\left\langle Y(A_3)/Y(A_2)\right\rangle$,
as function of the
temperature $T_9^\mathrm{f}(t_0)$ for a power-law decay of the
temperature during the second, slow expansion phase (blue curve)
compared to the dependence of this quantity on
$T_9^\mathrm{f}(t > t_0) = T_9^0$ in the case of constant asymptotic
temperature (red line, see also left panel of Fig.~\ref{sn_t9_nn}).
The other wind
parameters are again the same as in Figs.~\ref{nn_t145}--\ref{ynys_tf}
($s=145$, $Y_\indis{e}=0.42$, $\tau_\indis{dyn}=5\,$ms). The right panel
shows the distribution of cadmium isotopes at time $t = 0.1\,$s for the
values of $T_9^\mathrm{f}(t_0)$ listed in the inset; the neutron number
densities at $t = 0.1\,$s are given, too (cf.\ Fig.~\ref{ya_t145}).
The plot should be compared with the right panel of Fig.~\ref{sn_t9_nn}.
 }
  \label{sn_t9_nn2}
\end{figure*}

The left panel of Fig.~\ref{ynys_tf} displays the evolution of the
neutron-to-seed ratios $Y_\mathrm{n}/Y_\mathrm{seed}$ that correspond
to the neutron number densities of Fig.~\ref{ya_t145}. With an entropy
of $s = 145$ and an exponential timescale $\tau_{\mathrm{dyn}} = 5\,$ms
the values of $Y_\mathrm{n}/Y_\mathrm{seed}$ at the time when the
r-process has made the second abundance peak (at $t \approx t_0$) are
around 20 for $T_9^\mathrm{f}(t_0) = 0.4$--1.4. As discussed in
Sect.~\ref{sec:input}, this is a slightly insufficient number of
neutrons per seed nucleus to  create a third peak with exactly the
observed yields. The small underabundance of the third peak can also be
seen in the right panel of Fig.~\ref{ynys_tf}, where the height of the
platinum peak relative to the tellurium peak is given as function of
time. Different from the case of constant asymptotic temperature (see
Fig.~\ref{ynys_tc}), $\left\langle Y(A_3)/Y(A_2)\right\rangle$ is
nearly the same for $T_9^\mathrm{f}(t_0) = 1.0$ and 1.4, and is only
slightly reduced for $T_9^\mathrm{f}(t_0) = 0.4$. For even lower
transition temperatures the height of the third peak drops steeply.
This is evident from the blue line in the left panel of
Fig.~\ref{sn_t9_nn2} and is a consequence   of a faster decrease of the
free neutron density (compare the left panels in Figs.~\ref{nn_t145}
and \ref{ya_t145}), which for $T_9^\mathrm{f}(t_0)<0.4$ falls quickly
to a value of $n_\mathrm{n}\sim 10^{22}\,$cm$^{-3}$, below which the
equilibrium of n-captures and $\beta$-decays
is shifted to a region of slow beta-decay
rates, leading to reduction of the third peak.

The left panel of Fig.~\ref{sn_t9_nn2} provides a direct comparison
of the variation of $\left\langle Y(A_3)/Y(A_2)\right\rangle$ with
the transition temperature between the exponential first expansion
phase to the second phase of either constant temperature or power-law
temperature decrease. Above a transition temperature $T_9^\mathrm{f}(t_0)
\sim 0.4$ the behavior in both cases is dramatically different. While in
the case of constant asymptotic temperature the relative height of the
third peak decreases with higher values of $T_9^0$ (see Fig.~\ref{sn_t9_nn}
and Sect.~\ref{sec:constT}), a power-law decay of the temperature in
the second expansion phase leads to a prominent platinum peak for all
values of $T_9^\mathrm{f}(t_0)$ between 0.5 and 1.4. The reason is
again the large sensitivity of the photodisintegration reactions to
the late-time behavior of the temperature. Because of the power-law decline
the temperature drops within milliseconds to values where a
(n,$\gamma$)-($\gamma$,n) equilibrium is no longer possible but is replaced
by a quasi-equilibrium between (n,$\gamma$)-reactions and $\beta$-decays.
The nucleosynthesis at these conditions resembles the n-process of
Blake \& Schramm (1976), but there it was discussed to occur because of a
decrease of the neutron density below $10^{18}$, while here it happens
because of a decrease of the temperature and an associated strong
reduction of the ($\gamma$,n)-rates.
It should be noted that the prominent odd-even effect in the isotope
distributions in the right panel of Fig.~\ref{sn_t9_nn} for
$T_9^0 = 1.4$ and 0.4
has practically disappeared in Fig.~\ref{sn_t9_nn2}, where the
isotope distributions for the same values of the transition
temperature are much smoother.

We point out here that the r-processing of heavy nuclei
through a quasi-equilibrium of (n,$\gamma$)-reactions and
$\beta$-decays at conditions where photodisintegrations are
practically unimportant was recently also discussed by Wanajo
(2007), who coined the term ``cold r-process''\footnote{Different
from us, Wanajo (2007) used neutrino-driven wind
trajectories obtained as solutions of the steady-state wind
equations, and assumed the transition to a constant freeze-out
temperature $T_\mathrm{f}$ at some radius. As argued in
Sect.~\ref{sec:outflow}, such a more realistic description of the
early outflow dynamics in combination with constant conditions at
late times shares some basic features with the simple
two-stage expansion behavior considered in our work.}.
We prefer to call it ``r$\beta$-process'',
because this name is conform with the denotation of other processes
(r-process, rp-process, $\nu$p-process,...) and reflects the essential
aspect that characterizes this variant of the rapid neutron-capture
process.

\begin{figure}[tpb!]
 \begin{center}
 \includegraphics*[width=8cm]{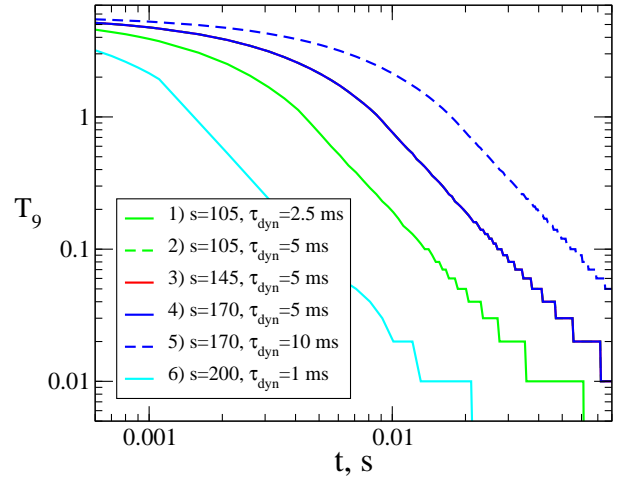}
 \end{center}
  \caption{Time evolution of the temperature for the different
  expansion histories considered in Fig.~\ref{nn_ya_all}. All have 
  an exponential first 
  expansion phase and second stage of power-law decline. Note 
  that the lines coincide for cases with different entropies $s$ but 
  the same expansion timescale $\tau_{\mathrm{dyn}}$.}
  \label{nn_t3}
\end{figure}

\begin{figure*}[tpb!]
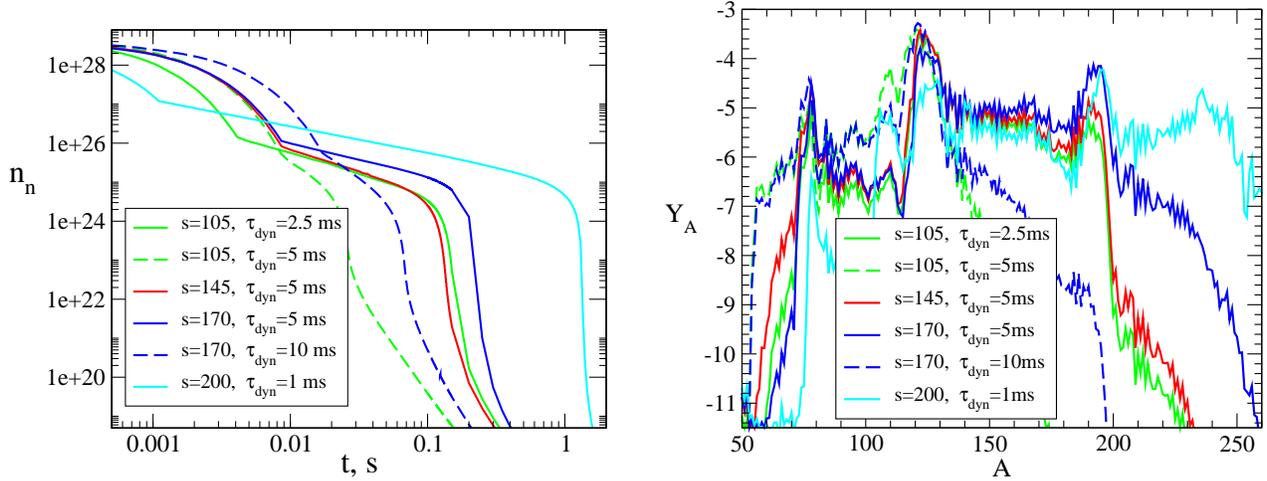

\begin{center}
\includegraphics*[width=8cm]{ff_t5nn2.eps}
\hspace{0.5cm}
\includegraphics*[width=8cm]{ff_t5ya.eps}
\end{center}
\caption{Time evolution of the free neutron density ({\em left}) and
final abundance distributions of r-process nuclei ({\em right}) for
outflows with different assumed entropies $s$ and exponential
expansion timescales $\tau_\mathrm{dyn}$ as listed in the inset.
Different entropy values are associated with different line colors,
a variation of the expansion timescale with dashed lines. In all
cases we assumed $Y_\mathrm{e} = 0.42$ and a transition temperature
of $T_9^\mathrm{f}(t_0) = 1$ between the exponential first expansion
phase and the second phase of power-law temperature decline. The 
corresponding temperature evolution is plotted in Fig.~\ref{nn_t3}.
The cyan line denotes the most extreme considered case with an
entropy of $s = 200$ and a dynamical timescale of $\tau_\mathrm{dyn}
= 1\,$ms. It serves for demonstrating the influence of fission
cycling (see also Fig.~\ref{a3a2s200}). }
 \label{nn_ya_all}
\end{figure*}

\subsection{Variations of wind parameters in the exponential phase}

In this section we will discuss the sensitivity of our nucleosynthesis
results to variations of the characteristic outflow conditions like
exponential expansion timescale, entropy, and electron fraction.
Table~\ref{tab:parameters} lists corresponding parameter values for
some of the considered outflows: $\tau_\mathrm{dyn}$ is the exponential
expansion timescale\footnote{We stress that comparing our expansion
timescale with those given in other papers requires some caution.
Otsuki et al.\ (2000),
Sumiyoshi et al.\ (2000), and Terasawa et al.\ (2002) defined the
expansion timescale as e-folding time at $T = 0.5\,$MeV, which is
compatible with our use. In contrast,
Wanajo et al.\ (2001, 2002) defined it as the cooling time
from $T = 0.5\,$MeV to $T = 0.2\,$MeV, which is about 10\% smaller.},
$T_9^\mathrm{f}(t_0)$ and $t_0$ the temperature and
time at which the transition occurs from the exponential first expansion
phase to the second phase with power-law decline of temperature and
density (the temperature evolution for different considered cases
is displayed in Fig.~\ref{nn_t3}), 
$v_\mathrm{ini}=R_\mathrm{ini}/\tau_\indis{dyn}$ is the initial
outflow velocity at an assumed initial radius $R_\mathrm{ini} = 10\,$km,
$v_0 = v_\mathrm{ini}\exp(t_0/\tau_\indis{dyn})$ is the outflow
velocity at transition time $t_0$, $\left\langle A\right\rangle$ is
the average mass number of nucleosynthesis yields, and
$\left\langle Y(A_3)/Y(A_2)\right\rangle$ the ratio of the height
of the third abundance peak relative to the second as defined in
Sect.~\ref{sec:constT}.


As discussed in Sect.~\ref{sec:input}, the strength of the r-process
is determined by the abundance of seed nuclei that is able to form,
and by the number of remaining free neutrons when the r-processing sets
in at $T_9 \approx 2$--3. For the homologous outflows considered there
we saw that a larger dynamical timescale increases the efficiency
of seed formation at the expense of free neutrons, and for a very
short dynamical timescale the expansion can be so rapid that
$\alpha$ particles and neutrons do not have any time to assemble to very
heavy elements. In both cases a strong third abundance peak cannot
develop. A smaller value of the outflow entropy also allows neutrons,
protons, and $\alpha$-particles to recombine to nuclei more efficiently
and therefore lower entropies have a similar effect as a slow expansion.

For a given value of the entropy, significant production of high-mass
elements therefore on the one hand requires that the expansion timescale
is sufficiently long, corresponding to a critical lower limit
$Y_\indis{s}^\mathrm{cr}$ of the seed abundance
at the start of the r-processing.
On the other hand, the expansion must be enough
fast because otherwise the seed production
exceeds $Y_\indis{s}^\mathrm{cr}$ too much, and the platinum
peak cannot be formed because of a disfavorably low neutron-to-seed
ratio. Our systematic runs for homologous outflows
showed (see Fig.~\ref{ynys_t9}) that with an
entropy of $s = 145$ and an electron fraction of $Y_\mathrm{e} = 0.42$
only
exponential timescales $\tau_\mathrm{dyn}$ of less than about 10$\,$ms
lead to a neutron number fraction of $Y_\indis{n} \ga 0.05$ and thus
to neutron-to-seed ratios around 20 at the beginning of the r-process.
Only then the nuclear flow has a chance to go beyond a mass number of
130 and to reach the range of $A \sim 196$,
although the third peak may still be significantly
underabundant compared to solar values (see below).
The critical limit for the seed abundance turned out to be
$Y_\indis{s}^\mathrm{cr} \sim 0.003$ (cf.\ Sect.~\ref{sec:input}).
Of course, this number depends on $s$ and $Y_\mathrm{e}$, and the
homologous expansion timescale that enables third peak
formation is shorter for lower entropy values.

Let us now discuss the case in which
the homologous outflow with its exponential density and
temperature decline is replaced by a slower power-law
temperature decay during the late expansion stage, i.e.\ after
the freeze-out of charged-particle reactions. This makes strong
r-processing up to the platinum peak possible for a much wider
range of dynamical timescales $\tau_\mathrm{dyn}$ than in the case
of purely homologous evolution, practically for all values below
some upper limit. This can
be seen in Fig.~\ref{nn_ya_all}, which displays the time-evolution
of the neutron number density and the final abundance distribution
for different choices of entropies and dynamical timescales.
The second stage of power-law temperature decline is assumed to
set in at $t = t_0$ with a transition temperature of
$T_9^\mathrm{f}(t_0) = 1$. A pronounced platinum peak develops
for $s = 105$ if $\tau_\mathrm{dyn} \la 2.5\,$ms,
for $s = 145$ this needs $\tau_\mathrm{dyn} \la 5\,$ms, and for
$s \approx 170$--200 it requires $\tau_\mathrm{dyn} \la 10\,$ms.
An extreme case with $s = 200$ and $\tau_\mathrm{dyn} = 1\,$ms
leads to fission cycling and demonstrates that even for very
fast homologous expansion during the first stage
the slower evolution in the second phase
allows all neutrons to be captured into heavy nuclei.
The dashed lines belong to cases where the expansion in the
homologous phase is too slow for third peak formation.

Figure~\ref{nn_ya_all} demonstrates that for a wide range of
timescale-entropy combinations in the considered intervals,
$2.5\,\mathrm{ms} \la \tau_\mathrm{dyn}\la 10\,$ms and $100 \la s
\la 200$, heavy r-process elements up to the third abundance peak
and beyond can be produced. Inspecting our results we find that a
very strong platinum peak appears for conditions that roughly
fulfill the relation $s \ga 10(\tau_\mathrm{dyn} + 10)$.

\begin{figure*}[tpb!]
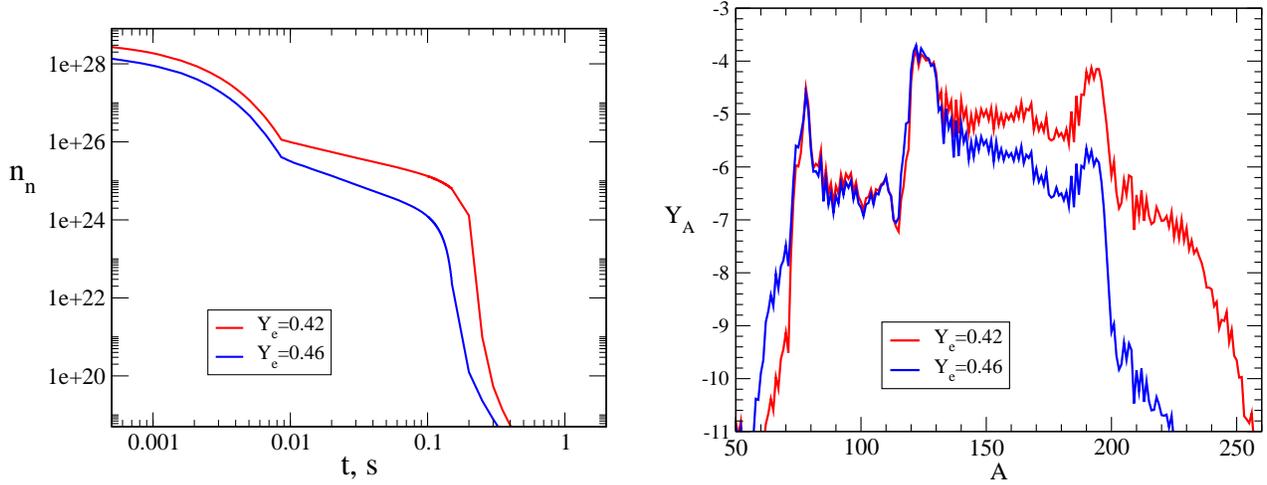

\begin{center}
\includegraphics*[width=8cm]{nf170t5nn.eps}
\hspace{0.5cm}
\includegraphics*[width=8cm]{nf170t5ya.eps}
\end{center}
\caption{Time-evolution of the free neutron number
density ({\em left}) and final abundance distribution of r-process nuclei
({\em right}) for $Y_\mathrm{e} = 0.42$ (red lines) compared to
$Y_\mathrm{e} = 0.46$ (blue lines). In both cases we used $s = 170$,
$\tau_\mathrm{dyn} = 5\,$ms, and a transition temperature of
$T_9^\mathrm{f}(t_0) = 1$ between the exponential first expansion phase
and the second phase of power-law temperature decline.}
\label{ye42_46_170}
\end{figure*}

In Fig.~\ref{ye42_46_170} the sensitivity of the r-processing
to variations of $Y_\indis{e}$ is shown in the case with $s = 170$ and
$\tau_\mathrm{dyn} = 5\,$ms. In the left panel one can see that a
change of $Y_\indis{e}$ from 0.42 to 0.46 leads to a reduction of
$n_\indis{n}$ at the beginning of the r-process and a more rapid
exhaustion of free neutrons. The yields beyond the
second peak are correspondingly lower, with a growing
discrepancy at higher mass numbers $A$ (right panel of
Fig.~\ref{ye42_46_170}. While for $Y_\indis{e} = 0.42$ the
platinum peak is slighly overabundant compared to observations
(see below), increasing $Y_\indis{e}$ by about 10\% to 0.46
is enough to lead to a significant underproduction.
For the considered short dynamical timescale this can be
compensated by a roughly 30\% higher value of the outflow entropy.

\begin{figure*}[tpb!]
\begin{center}
\includegraphics*[width=8cm]{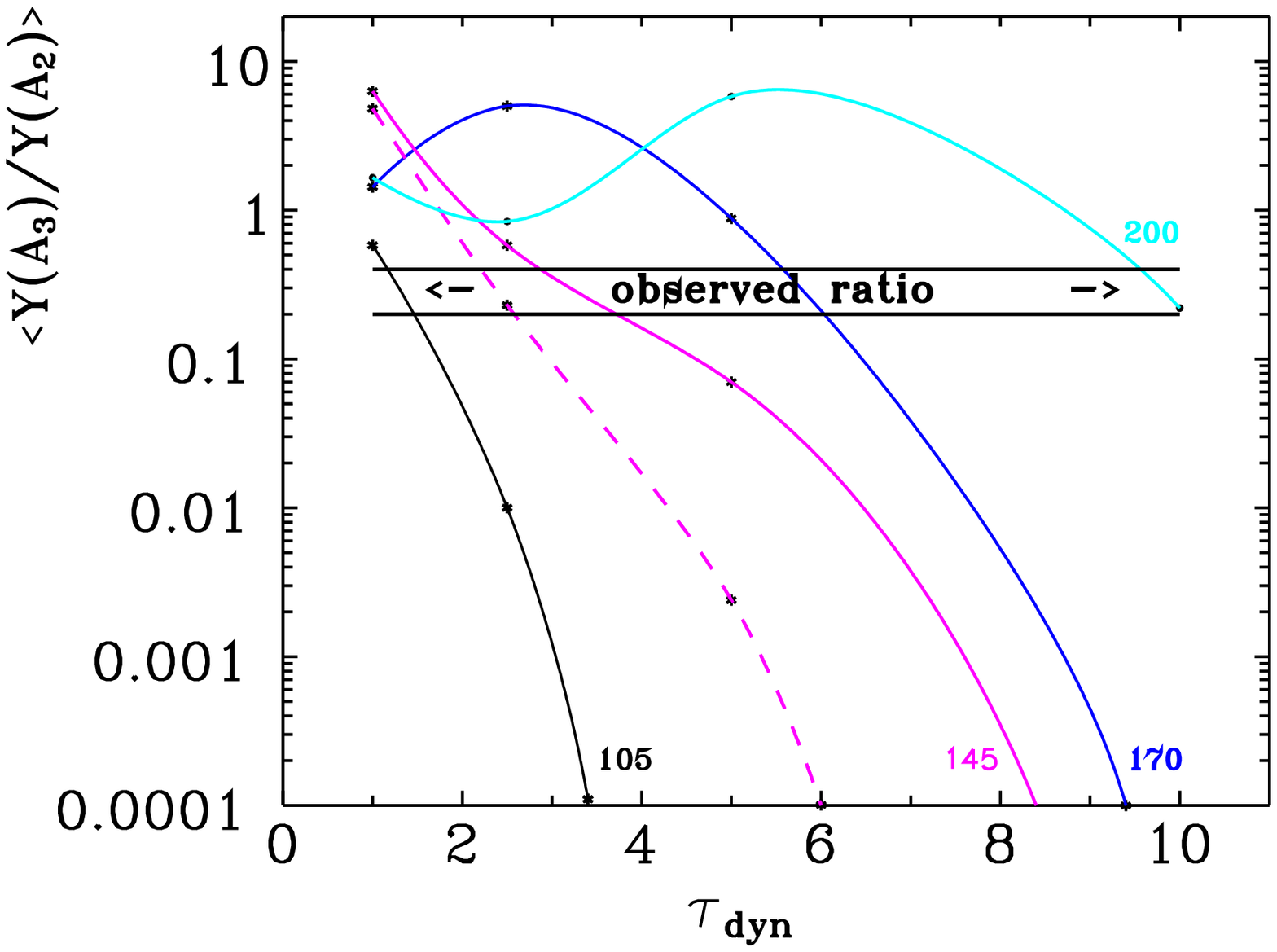}
\hspace{0.5cm}
\includegraphics*[width=8cm]{a3a2s200t1_1f2.eps}
\end{center}
\caption{{\em Left:} Height of the third abundance peak relative to the
second as function of the exponential expansion timescale
$\tau_\mathrm{dyn}$, for different values of the entropy (labels at
curves). The solid lines correspond to outflows with a power-law
temperature decay following the exponential expansion, while the dashed
line shows for comparison the results for $s = 145$ and constant
temperature $T_9^\mathrm{f}(t > t_0) = T_9^0$ during the second, slow
expansion phase. In all cases the outflow was assumed to have an
electron fraction of $Y_\mathrm{e} = 0.42$ and the transition
temperature was taken to be $T_9^\mathrm{f}(t_0) = 1$. The band between
the two bold horizontal lines marks the observational range. {\em
Right:} Time evolution of the ratio of the third abundance peak to the
second peak and of the second and third abundance peaks, $Y(A_2)$ and
$Y(A_3)$ (according to the definition in Eq.~\ref{eq:peak_heights}),
normalized to the total amounts of seed nuclei, $Y_\mathrm{seed}$, at any
given time $t$ (dashed lines) and to the seed abundance at the end of
the evolution at $t = 2\,$s (solid lines), respectively; the different
line styles and colors are labelled in the inset box. Note that we
consider all nuclei with $Z > 2$ as seeds. The results were computed
for $s = 200$, $\tau_\mathrm{dyn} = 1\,$ms, $Y_\mathrm{e} = 0.42$, and
$T_9^\mathrm{f}(t_0) = 1$.
The waves visible in the time evolution of the different
quantities reflect the effects of fission cycling in an environment
with a very high free neutron density, which is the case for the
extreme parameters of this run. }
  \label{a3a2s200}
\end{figure*}

In Fig.~\ref{a3a2s200} we provide an overview of the platinum peak
formation in dependence of the entropy and dynamical timescale of
the model outflows, using $Y_\mathrm{e} = 0.42$ and assuming a transition
temperature of $T_9^\mathrm{f}(t_0) = 1$ between the exponential first
cooling phase and the second phase with power-law decrease of
temperature and density. The left panel confirms what we
described above: for higher values of the entropy the $A \sim 196$
peak can be assembled for an increasingly wider range of expansion
timescales. The bold horizontal lines in the left panel mark the
observational band for the abundance
ratio $\left\langle Y(A_3)/Y(A_2)\right\rangle$.
The third peak relative to the second tends to become overabundant
compared to the observations when the expansion timescale is low,
whereas it remains too weak for long expansion timescales.

The dashed line in the left panel of Fig.~\ref{a3a2s200}
corresponds to runs with $s = 145$
where a constant density and temperature
were adopted during the second phase instead of the power-law
behavior. The differences between the dashed and solid lines
for the same entropy are large, in particular for
longer dynamical timescales. This demonstrates again the
importance of the late-time behavior of the outflow. This
importance, however, is significantly reduced when the
transition temperature $T_9^\mathrm{f}(t_0)$ is chosen to be
near 0.4 (see the left panel of Fig.~\ref{sn_t9_nn2}).

For entropies near 200 or higher fission cycling can occur.
The right panel of Fig.~\ref{a3a2s200} shows the time evolution of the
second and third abundance peaks and of the ratio between both
for a case with $s = 200$, $\tau_\mathrm{dyn} = 1\,$ms, and
$Y_\mathrm{e} = 0.42$. The periodic changes of the abundances and
of $\left\langle Y(A_3)/Y(A_2)\right\rangle$ reflect the repeated
propagation of nucleosynthesis wave through the
transuranium region via fission cycling (the number of nuclear
species is increased roughly by a factor of three due to fission),
similar to what was observed by Panov et al.\ (2003),
Goriely et al.\ (2005), and Beun et al.\ (2008).
The overabundance of the platinum peak
in this case depends significantly on the nuclear data
(\cite{pft01}) and should be explored separately on the basis of new
fission data calculations (Panov et al.\ 2005).

\section{Discussion and conclusions}

Focussing on the high-entropy neutrino-driven wind environment, we
investigated the sensitivity of r-process nucleosynthesis
to the asymptotic
low-temperature behavior that is assumed to follow an approximately
homologous early expansion of very rapidly accelerated
outflows. On the one hand we compared the continuously homologously
evolving case to a situation where the temperature and density reach
a lower limit and remain constant later on.
The latter setting is equivalent to the choice
of a freeze-out temperature for supersonic wind solutions
by Wanajo et al.\ (2002) and Wanajo (2007), and it is similar to
the use of a constant value for the
outer boundary pressure of subsonic breeze solutions as adopted by
Sumiyoshi et al.\ (2000), Terasawa et al.\ (2002), and Wanajo et al.\
(2001). Breeze solutions, however, are characterized by the very important
feature that they imply a causal connection between the physical
conditions at the neutron star and at the outer boundary, i.e. the
mass-loss rate and asymptotic temperature are not independent, whereas
supersonic outflows with a limiting temperature are not subject to
such a constraint. The constant conditions at late times correspond to
a situation where the homologous outflow is asymptotically decelerated
to very small (zero) velocity. This can be considered as a reasonably good
representation of the situation when a supersonic wind is slowed down
in a termination shock at colliding with the preceding supernova
ejecta. On the other hand we investigated a case where the
velocity was assumed to make a transition from the linear radial growth
of the homologous phase to an asymptotically constant, nonvanishing value.
In this case the density was assumed to decrease with time like $t^{-2}$
and the temperature (for constant radiation entropy) like $t^{-2/3}$.
This is supposed to be a model for the situation when a supersonic
wind reaches a second stage of slower acceleration at late times.

We have found that for given and constant wind radiation entropy
(we explored $s\sim 100$--200$\,k_\mathrm{B}$ per nucleon),
electron-to-baryon ratio ($Y_\mathrm{e}\sim 0.42$--0.46), and
(sufficiently small)
wind expansion timescale ($\tau_\mathrm{dyn}\sim 1$--10$\,$ms),
a strong r-process with production of the third
abundance peak depends not only on the value of the transition temperature
between the two expansion phases,
but also on the evolution of temperature and density during the second,
slower stage. In the case of a constant
asymptotic temperature and density, the formation of
the platinum peak is enabled when the asymptotic temperature value
is moderately high ($T_9^0\sim 0.2$--0.8). When leaving this range
towards lower or higher asymptotic temperatures, the
possibility of third-peak formation strongly decreases. This behavior
was also seen by Wanajo et al.\ (2002) and Terasawa et al.\ (2002)
for, as they called them, ``freeze-out temperatures'' or ``boundary
temperatures'' above $T_9^0\sim 0.4$, but the variation with even lower
temperatures remained unexplored in both works.

In our second considered case with power-law decline of $T(t)$ and
$\rho(t)$, the strength of the $A\sim 195$ abundance peak compared
to the $A\sim 130$ peak turned out to be relatively independent of
the transition temperature $T^{\mathrm{f}}(t_0)$ in the interval
between about $3\times 10^8\,$K and $1.4\times 10^9\,$K. For
lower transition temperatures the third-peak formation decreases
steeply and behaves
similar to the case with constant asymptotic temperature and density.
This means that the appearance and disappearance of a prominent
platinum peak when the constant asymptotic temperature
is lowered from $T_9^0\sim 1$ to $T_9^0\sim 0.1$, is not observed in
the case of a power-law decline of temperature and density during
the late wind evolution. Interestingly, both considered cases of
late-time expansion behavior also lead to distinctive differences
in the abundances of neighboring isotopes of r-process elements.
For the power-law time-dependence not only more neutron-rich
isotopes are formed but the isotopic distribution is also smoother.

Strong r-processing naturally requires a sufficiently fast
expansion in the homologous phase so that charged-particle
reactions freeze out before excessive seed production occurs and
high neutron-to-seed ratios cannot be reached.
Moreover, a relatively rapid temperature
decline during the second, slower expansion phase, and at the same
time a persistence of sufficiently high neutron number densities
($n_\mathrm{n}\ga 10^{24}\,$cm$^{-3}$) are favorable for
driving the nuclear flow beyond the second abundance peak. For fixed
radiation entropy $s$ as assumed in our models, temperature and
density are coupled by the relation $s \propto \rho/T^3 =
{\mathrm{const}}$. In this case the asymptotic power-law decrease of
both quantities enables third peak formation for a wider range of
transition temperatures.

Terasawa et al.\ (2002) argued that lower asymptotic temperatures
reduce the charged-particle reactions, which leads to less seed production
and a higher neutron-to-seed ratio, thus causing a better agreement
of the r-process yields with the solar abundances. For our calculations,
which are based on the idealized two-stage model of the wind
expansion, this argument cannot be made. Long before the corresponding
asymptotic temperatures (between $T_9 = 0.1$ and $T_9 = 1.4$) are
reached, namely already above $T_9 \approx 2$, charged-particle
reactions become inefficient because of the impenetrability of the
nuclear Coulomb barrier for low-energy thermal protons. Instead,
the described nucleosynthesis results can be understood by the density
and temperature dependence of neutron-captures and nuclear
photodisintegration reactions.

Since the ($\gamma$,n)-rates decrease steeply with falling temperature,
their role diminishes with a lower transition temperature. For
sufficiently high neutron densities the
still rapid neutron captures force the r-process path towards the
neutron-drip line, where the $\beta$-decay rates are large. Instead of
going through a ($\gamma$,n)-(n,$\gamma$) equilibrium, the r-processing
proceeds now as a quasi-equilibrium of (n,$\gamma$) reactions and
$\beta$-decays, which suggests the term ``r$\beta$-process''. This
sitation in outflows that expand supersonically and cool quickly to
a few $10^8\,$K was recently also discussed by Wanajo (2007), who
named it ``cold r-process''. The formation of the third abundance peak
is possible in this situation, provided the neutron-to-seed ratio is
large enough, because the $\beta$-decays are fast and allow for
a quick assembling of heavy nuclei. In contrast, if the exponential
expansion makes a transition to a constant temperature that is high,
($\gamma$,n)-reactions drive the r-process path towards the valley of
stability where the $\beta$-decay rates are small and the r-processing
therefore slows down. In this case the free neutrons are used up in
forming the second abundance peak before any significant third
maximum can build up. On the other hand, the power-law temperature
decline in the slow expansion phase leads to an efficient r$\beta$-process
for a wide range of transition temperatures $T_9^{\mathrm{f}}(t_0)$.
Because the strength of
the third abundance peak then depends only on the neutron density, it
becomes relatively insensitive to the value of $T_9^{\mathrm{f}}(t_0)$
between about 0.3 and 1.4. If the transition temperature is even lower,
the free neutron density during the slow expansion phase of the
outflow is not large enough any more to support a strong
r-processing.

In previous investigations (Terasawa et al.\ 2001, 2002; Wanajo et al.\
2002) the complex influence of the late-time outflow behavior was not
explored, but it matters when the robustness
of abundance yields to variations in the neutrino-driven wind models
is discussed, especially in situations where high entropies, low
temperatures, and high neutron densities lead to a change of
the r-process to an r$\beta$-process.

In summary, we therefore conclude that the detailed cooling
behavior during the late expansion of supernova outflows can
have important consequences for r-process nucleosynthesis.
In particular, a
slow expansion phase with decreasing temperature and density
following a rapid, supersonic initial outflow expansion, is more
favorable for a strong r-process than the constant boundary
conditions assumed previously (e.g., Wanajo et al.\ 2001, 2002;
Terasawa et al.\ 2001, 2002).
Despite differences in the initial neutron
number density by more than two orders of magnitude, such
conditions can lead to a fairly robust production of the platinum
peak for a wider range of transition temperatures (at least
for the cases of very fast early expansion and modest values of
entropy and neutron excess as considered in our study). This is
caused by a kind of self-regulation of the quasi-equilibrium
of ($\gamma$,n)-reactions and $\beta$-decays in the
r$\beta$-process, which leads to roughly the same number of
neutron captures per seed nucleus, independent of whether the
process proceeds quickly for high neutron densities (in which
case the r-process path is closer to the neutron-drip line) or
slower in the case of low neutron densities. This might give a
hint for the outflow dynamics of neutrino-driven wind ejecta that
can enable the uniform production of heavy r-process elements
suggested by the abundance patterns observed in metal-poor
stars (see e.g., Cowan \& Sneden 2006).

For the explored range of parameter values, the r-processing
can last 150--200$\,$ms or even longer. This duration
is the lower time limit for the formation of heavy elements
from iron-group seeds up to the platinum peak. For entropies
$s$ between 100 and 200$\,k_\mathrm{B}$ per nucleon,
a solar-like production of the heaviest elements can occur
only in the case that the outflow expansion decelerates
and a transition from the initial homologous phase
to a slower later stage of expansion occurs. In addition, a short
or very short exponential timescale ($\tau_{\mathrm{dyn}}\sim
10\,$ms for $s = 200$ and $\tau_{\mathrm{dyn}}\sim 1\,$ms for
$s = 100$) is needed during the homologous expansion, if moderately
neutron-rich conditions ($Y_\mathrm{e} = 0.42$) are considered.
Of course, if the neutron excess decreases, the formation of the
third abundance peak requires a higher entropy. These parameter
constraints are
rather similar to those shown in Fig.~10 of Hoffman et al.\ (1997).
This can be understood from the fact that supersonic
solutions of the neutrino-driven wind equations as those employed by
Hoffman et al.\ (1997) show approximately
homologous expansion ($v \propto r$) only up to some maximum radius,
above which the velocity continues to grow significantly less
rapidly. Although there is no deceleration as it is caused by a wind
termination shock (see Arcones et al.\ 2007), the slow-down of the
acceleration of supersonic winds has still some basic similarity to
the transition from a rapid first expansion phase to a slower second
stage as assumed in our simplified outflow models.

\section*{Acknowledgements}

\begin{acknowledgements}
We are grateful to an anonymous referee for very valuable suggestions
to improve our presentation.
This work was in part supported by the SNF project
No.~IB7320-110996 and by the Deutsche Forschungsgemeinschaft
through the Transregional Collaborative Research Centers SFB/TR~27
``Neutrinos and Beyond'' and SFB/TR~7 ``Gravitational Wave Astronomy'', 
and through the Cluster of Excellence EXC~153 ``Origin and Structure 
of the Universe'' ({\tt http://www.universe-cluster.de}).
\end{acknowledgements}



\begin{thebibliography}{}


\bibitem[2007]{almu07} Arcones, A., Janka, H.-Th., \& Scheck, L. 2007, A\&A,
467, 1227

\bibitem[2007]{arno07} Arnould, M., Goriely, S., \& Takahashi, K. 2007,
Physics Reports, 450, 97

\bibitem[2008]{beun08} Beun, J., McLaughlin, G.C., Surman, R., \&
Hix, W.R. 2008, Phys.\ Rev.\ C, 77, 035804


\bibitem[1979]{BKChec79 } Bisnovatyi-Kogan, G. S. \& Chechetkin, V. M. 1979,
 Sov. Uspehi Phys. Nauk,
          127, 263

\bibitem{Shr76} Blake, J. B., \& Schramm, D. N.
1976,            Astrophysical Journal,  209, 846


\bibitem[1996]{bp96} Blinnikov, S. I., \& Panov, I. V. 1996,
           Astronomy Letters,   22, 39

\bibitem[2006]{bur06} Buras, R., Janka, H.-Th., Rampp, M., \& Kifonidis,
K. A\&A, 2006, 457, 281  


\bibitem[1957]{bbfh57}    Burbidge, G. R., Burbidge, E. M., Fowler, W. A., \& Hoyle, F. 1957,
         Rev. Mod. Phys.,  29, 547


\bibitem[1995]{burrows95} Burrows, A., Hayes, J., \& Fryxell, B. A., 1995,
  ApJ, 450, 830



\bibitem[2002]{camer03} Cameron, A. G. W. 2003,  ApJ,  587, 327


\bibitem[1983]{cam83} Cameron, A. G. W., Cowan, J. J., \& Truran, J. W.
   1983,  Astrophysics and Space Science, 91, 235


\bibitem[1997]{carful97}
    Cardall, C. Y., Fuller, G. M. 1997,     ApJ,  486,  L111


\bibitem[2006]{cowan06} Cowan, J. J. \& Sneden, C. 2006, Nature, 440, 1151


\bibitem[1991]{ctt91} Cowan, J. J., Thielemann, F.-K., \& Truran, J. W. 1991,
         Phys. Rep.,  208, 267


\bibitem[Freiburghaus et al. (1999)]{frt99}  Freiburghaus, C.,  Rosswog,
S.,\&  Thielemann, F.-K. 1999,  ApJ,   525, L121


\bibitem[1971]{gear71}   Gear, C. W., 1971, Numerical Initial Value
Problems in Ordinary  Differential Equations
 (Englewood Cliffs, New Jersey: Prentice-Hall)

\bibitem[1996]{gor96} Goriely, S. \& Arnould, M. 1996, A\&A, 312, 327


\bibitem[2005]{gor05} Goriely, S., Demetriou, P., Janka, H.-Th., 
Pearson, J. M., \&
Samyn, M.  2005, Nuclear Physics A,  758  587c 


\bibitem{wfh78}  Hillebrandt, W. 1978, Space Sci. Rev.,   21, 639



\bibitem{Hoff97} Hoffman, R. D., Woosley, S. E., \& Qian, Y.-Z. 1997,  ApJ,
482,  951


\bibitem{Imsh92}  Imshennik, V. S. 1992,   Astronomy Letters,  18, 194

\bibitem{Imsh02} Imshennik, V. S., Litvinova I. Yu. 2006. Atomic Nuclei, 69,
660.

\bibitem[1995a]{jan95a} Janka, H.-Th.,  \& M\"uller, E.    1995,
  Physics Reports,   256,  135  

\bibitem[1995b]{jan95b} Janka, H.-Th.,  \& M\"uller, E.    1995,
  ApJ, 448, L109


\bibitem{klk93} Kratz, K.-L., Bitouzet, J.-P.,  Thielemann, F.-K.,
 M\"oller, P., \& Pfeiffer, B. 1993, ApJ,
     403, 216

\bibitem{kur07} Kuroda, T., Wanajo, S., \& Nomoto, K. 2008, ApJ, 672, 1068

\bibitem{LatShr74}  Lattimer, J., M., \&  Schramm, D. N. 1974,
        ApJ,   192, L145


\bibitem[Lattimer \& Schramm, 1976]{LatSch76} Lattimer, J. M., Schramm, D. N., 1976,
         ApJ,  210, 549

\bibitem[1992]{meyer92} Meyer, B. S., Mathews, G. J., Howard, W. M., Woosley, S. E.,
    \&   Hoffman, R. D., 1992,  ApJ,   399, 656.

\bibitem[1995]{moell95} M\"oller, P., Nix, J. R., Myers, W. D., \& 
Swiatecki, W. J. 1995, Atomic Data Nucl.\ Data Tables, 59, 185


\bibitem{npb98} Nadyozhin, D. K., Panov, I. V., \&
            Blinnikov, S.I.  1998, A\&A,   335,   207.

\bibitem{nadyu04}   Nadyozhin, D. K., \& Yudin, A. V. 2004, Astronomy Letters,
  30, 697

\bibitem{otsu00} Otsuki, K., Tagoshi, H., Kajino, T., \& Wanajo, S. 2000,
 ApJ, 533, 424

\bibitem[Panov 2003]{pan_r3}  Panov, I. V. 2003, Astronomy Letters, 29, 163

\bibitem[Panov \& Chechetkin 2002]{pan02che}  Panov, I. V. \& Chechetkin, V. M.
 2002, Astronomy Letters, 28, 476

\bibitem[Panov \& Thielemann 2003]{Jap02np3}  Panov I. V. \& Thielemann F.-K.,
2003, Nucl. Phys. A, 718, 647



\bibitem[Panov et al.\ (2001a)]{pbt01} Panov, I. V., Blinnikov, S. I., \& Thielemann, F.-K. 2001a,
           Astronomy Letters,  27,  1



\bibitem[Panov et al.\ 2001b]{pft01}  Panov, I. V., Freiburghaus, C.,
\& Thielemann, F.-K. 2001b, Nucl. Phys. A,  688, 587


\bibitem[Panov et al.\ 2005]{panfis05}  Panov, I. V., Kolbe, E., Pfeiffer, B., Rauscher, Th.,
Kratz, K.-L., \& Thielemann, F.-K. 2005, Nuclear Physics A, 747, 633




\bibitem{QiWas00} Qian, Y.-Z. \& Wasserburg, G. J. 2000,
         Physics reports, 333, 77

\bibitem{qiwo96} Qian, Y.-Z. \& Woosley, S.E. 1996,  ApJ,   471, 331

\bibitem{RTK97} Rauscher, T., Thielemann, F.-K., \& Kratz,
K.-L. 1997, Nucl. Phys., A 621,  331c.

%
\bibitem{rath00} Rauscher, T. \& Thielemann, F.-K. 2000,
  Atomic Data Nucl. Data Tables,   75, 1


\bibitem{SysShr82} Symbalisty, E. M. D. \& Schramm, D. N.
1982,            Astrophysical Letters,   22, 143

\bibitem{SFC65} Seeger, P. A., Fowler, W. A., \& Clayton, D. D. 1965,
 ApJS,   11, 121


\bibitem{sumi00} Sumiyoshi, K., Suzuki, H., Otsuki, K., Terasawa, M.,
\&  Yamada Sh. 2000, Publ. Astron. Soc. Japan., 52, 601




\bibitem[1994]{twj94} Takahashi, K.,  Witti, J.,  \& Janka, H.-T. 1994,
        A\&A,   286, 857

\bibitem[1997]{taja97} Takahashi, K. \& Janka, H.-T. 1997,
  Proceedings of Int. Conf. {\em Origin of Matter and Evolution of 
  Galaxies in the Universe '96}.
  Atami, Japan, 18-20 January 1996, edited by T. Kajino, Y. Yoshii, 
  and S. Kubono (World Scientific, Singapore), p. 213 

\bibitem[2001]{tera01} Terasawa, M., Sumiyoshi, K., Kajino, T., Mathews, G. J., \&
  Tanihata, I. 2001,  ApJ, 562, 470

%
\bibitem{tera02} Terasawa, M., Sumiyoshi, K., Yamada, S., Suzuki, H., \& Kajino, T. 2002,  ApJ, 578, L137
%
%
\bibitem{thi87}
Thielemann, F.-K., Arnould, M. \& Truran, J.W. 1987, in {\em {A}dvances
in  {N}uclear {A}strophysics}, edited by E. Vangioni-Flam et~al.
(Editions fronti\`ere, Gif sur Yvette), p.\ 525
%
%
\bibitem{Thom01} Thompson, T. A., Burrows, A., \& Meyer, B. S. 2001,  ApJ, 562
 887


\bibitem[2004]{tomas04} Tom{\`a}s, R., Kachelrie\ss, M., Raffelt, G., Dighe, A., Janka, H.-T., \& Scheck, L.  2004,
 Journal of Cosmology and Astroparticle Physics,    09,   015       



\bibitem{Wana07} Wanajo, S. 2007, ApJ, 666, L77

\bibitem{WanaIshi06}  Wanajo, S., \& Ishimaru, Y. 2006, Nuclear Physics A,
777, 676


\bibitem{Wana01} Wanajo, S.,   Kajino, T., Mathews, G. J., \& Otsuki,
K. 2001,   ApJ,  554, 578

\bibitem{Wana02} Wanajo, S., Itoh, N., Ishimaru, Yu.,  Nozawa,
S., \&  Beers, T. C. 2002, ApJ, 577, 853

\bibitem{Wana04} Wanajo, S., Goriely, S., Samyn, M., \& Itoh, N. 2004,
ApJ, 606, 1057

\bibitem[1994]{wjt94}  Witti, J., Janka, H.-T., \& Takahashi, K. 1994, A\&A,  286, 841

\bibitem[1993]{witti93} Witti, J., Janka, H.-Th., Takahashi, K., \&
Hillebrandt, W. 1993,
      In Nuclei in the Cosmos-IX, Eds. F. K\"appeler, K. Wisshak
      (Bristol and Philadelphia: Inst. of Phys. Publ.),  p. 601

\bibitem{WoHof92} Woosley, S. E. \& Hoffman, R. D. 1992,  ApJ,
 395, 202

\bibitem{Woos94r}   Woosley, S. E.,  Wilson, J. R.,  Mathews, G. J.,
Hoffman, R. D., \& Meyer, B. S. 1994, ApJ,  433, 229


\end{thebibliography}
\end{document}